\documentclass[a4paper,fleqn]{cas-sc}

\usepackage[numbers,sort&compress]{natbib}

\usepackage{lineno}
\usepackage{amssymb}
\usepackage{amsmath}
\usepackage{amsfonts}
\usepackage{color}
\usepackage{soul}
\usepackage{ulem}
\usepackage{float}
\usepackage{caption}
\begin{document}
\let\WriteBookmarks\relax
\def\floatpagepagefraction{1}
\def\textpagefraction{.001}
\shorttitle{}
\shortauthors{Y. Ye et~al.}
\let\printorcid\relax
%\begin{frontmatter}

\title [mode = title]{Nonreciprocal parametric amplification of elastic waves in supersonic space-time modulated media}                      
%\tnotemark[1,2]

%\tnotetext[1]{This document is the results of the research
   %project funded by the National Science Foundation.}

%\tnotetext[2]{The second title footnote which is a longer text matter
  % to fill through the whole text width and overflow into
  % another line in the footnotes area of the first page.}

\author[a,b]{Yingrui Ye}[]
\author[c]{Chunxia Liu}[]
%\author[b]{Alessandro Marzani}[]
%\author[d]{Emanuele Riva}[]
\author[a]{Xiaopeng Wang}[]\cormark[1]\ead{xpwang@mail.xjtu.edu.cn}
\author[b]{Antonio Palermo}[]\cormark[1]\ead{antonio.palermo6@unibo.it}

\cortext[cor1]{Corresponding author}

\affiliation[a]{organization={Department of Mechanical Engineering, Xi'an Jiaotong University},%Department and Organization
           %% addressline={}, 
            city={Xi'an},
            postcode={710049}, 
          %%  state={},
            country={China}}
\affiliation[b]{organization={Department of Civil, Chemical, Environmental and Materials Engineering, University of Bologna},%Department and Organization
           %% addressline={}, 
            city={Bologna},
            postcode={40136}, 
          %%  state={},
            country={Italy}}            
\affiliation[c]{organization={College of Electrical Engineering and Automation, Anhui University},%Department and Organization
           %% addressline={}, 
            city={Hefei},
            postcode={230601}, 
          %%  state={},
            country={China}}
\affiliation[d]{organization={Department of Mechanical Engineering, Politecnico di Milano},%Department and Organization
           %% addressline={}, 
            city={Milano},
            postcode={20156}, 
          %%  state={},
            country={Italy}}

\begin{abstract}
Space-time modulated elastic media, whose material properties vary in both space and time, have attracted significant attention as a promising strategy for achieving nonreciprocal propagation of elastic waves. To date, most studies have focused on systems with subsonic modulation, where the phase velocity of the modulation wave is lower than that of the guided elastic waves. In contrast, wave propagation under supersonic modulation remains largely unexplored, and the associated nonreciprocal phenomena present an open challenge. In this work, we investigate the dynamic response of elastic longitudinal waves propagating through a spatially bounded medium subjected to supersonic modulation of its elastic properties. We show that supersonic modulation gives rise to directional wavenumber bandgaps in the dispersion diagram, characterized by vanishing wavenumbers. Using a theoretical framework based on mode-coupling theory, we reveal how supersonic modulation governs amplitude modulation and frequency conversion in wave transmission and reflection at spatial interfaces. This leads to nonreciprocal parametric amplification and frequency conversion, phenomena not previously reported in bounded elastic media. Remarkably, the observed parametric amplification is not driven by an imaginary component of the wavenumber, but instead arises from the interference between counter-propagating Floquet-Bloch modes at spatial interfaces. Numerical simulations corroborate the theoretical predictions and illustrate nonreciprocal amplification and frequency conversion effects. Our findings pave the way for the design of nonreciprocal elastic wave devices with advanced functionalities such as signal processing, energy amplification, and frequency conversion.
\end{abstract}

%\includegraphics{figs/cas-grabs.pdf}

%\item Directional wavenumber bandgaps are tailored under supersonic modulation.
%\item The scattering behavior of elastic waves in space-time modulated media is modeled.
%\item Nonreciprocal parametric amplification and frequency conversion are demonstrated.
%\item The amplification mechanism arises from interference between Floquet-Bloch modes.


\maketitle

%\begin{linenumbers}

\section{Introduction}
\label{sec1}

The study of nonreciprocal wave phenomena, where the receiver and source are not interchangeable, has garnered significant interest in both physics and engineering communities \citep{sounas2017non,nassar2020nonreciprocity,nagulu2020non,kutsaev2021up,wang2023mechanical,yang2024nonreciprocal}. This interest stems from the potential to develop devices with uncommon wave manipulation capabilities, such as acoustic rectifiers \citep{liang2010acoustic}, circulators \citep{fleury2014sound}, isolators \citep{zhang2024nonreciprocal}, optical, acoustic, and mechanical diodes \citep{wang2013optical,li2011tunable,jalvsic2023active}, unidirectional amplifiers \citep{koutserimpas2018nonreciprocal}, unidirectional frequency converters \citep{shen2023nonreciprocal,guo2023observation}, and topological insulators \citep{fleury2016floquet,tian2022experimental}. Nonreciprocity can be generally obtained by following four distinct strategies. The first approach is to introduce circulating flows or moving parts to bias the waveguide motion \citep{fleury2014sound,virtanen2024nonreciprocal}. The second one is to combine material nonlinearity and spatial asymmetry to break reciprocity \citep{lepri2011asymmetric,cotrufo2024passive}. The third one is to construct two different Chern insulators which give rise to topologically protected one-way edge states at their interfaces \citep{ma2019topological,wang2021topological_2}. The fourth way is to break the time-reversal symmetry in the absence of material dissipation by realizing space-time-varying constitutive properties, i.e., spatiotemporal modulation \citep{nassar2017modulated,galiffi2022photonics}.

Among the various approaches, spatiotemporal modulation is particularly attractive, as it can be practically implemented in conventional waveguide platforms, for example by incorporating switchable shunt circuits \citep{Trainiti2019time,Marconi2020experimental,tessier2023experimental}. Research on space-time periodic media dates back to the 1960s, when such systems were first explored in the context of electrical circuits \citep{cassedy1963dispersion, cassedy1967dispersion}. In the field of elasticity, interest in wave phenomena within spatiotemporally modulated systems has emerged more recently. Over the past decade, such systems have been explored both theoretically and experimentally, either through direct modulation of the host medium’s material properties \citep{trainiti2016non, nassar2017modulated, goldsberry2020nonreciprocal, Marconi2020experimental} or by using arrays of modulated resonators \citep{nassar2017non, chen2019nonreciprocal, wu2021non, palermo2020surface}. In these spatiotemporally modulated elastic systems, a series of remarkable wave propagation phenomena have been demonstrated, including nonreciprocal energy isolation and frequency conversion.

While most research has focused on dispersion relations and free wave propagation in infinite modulated elastic media, the effects of spatial interfaces and the associated wave propagation in bounded space-time modulated systems remain relatively unexplored. A few examples include theoretical investigations of interfaces between semi-infinite homogeneous and spatiotemporally modulated media \citep{yi2017frequency}, as well as studies on the reflection and transmission properties of elastic waves propagating through finite spatiotemporally modulated media bounded by two spatial interfaces \citep{yi2018reflection}. Still, most studies to date have focused on subsonic modulation, referring to scenarios in which the phase velocity of the pump wave is lower than that of the medium-supported wave. This emphasis stems from the assumption that stable system responses cannot be achieved under supersonic modulation \citep{cassedy1963dispersion, cassedy1967dispersion, nassar2017modulated}. Consequently, a comprehensive understanding of scattering phenomena and wave amplification at spatial interfaces in bounded, supersonically modulated media remains an open and important challenge.

In this work, we present a thorough investigation of the scattering behavior of elastic waves at spatial interfaces in a spatiotemporally modulated medium subjected to supersonic modulation. Specifically, we examine the propagation of elastic longitudinal waves through a finite, supersonically modulated medium bounded by two spatial interfaces. To predict the scattering coefficients in this spatially confined system, we adopt a theoretical framework based on mode-coupling theory. This model enables us to capture nonreciprocal parametric amplification induced by supersonic space-time modulation in elastic media. Further analysis of the Floquet-Bloch modes reveals that the amplification mechanism stems from interference between counter-propagating components. We validate the theoretical predictions with numerical simulations, which also highlight peculiar nonreciprocal wave phenomena, including parametric amplification and frequency conversion.

The paper is organized as follows. \textcolor{blue}{Section \ref{sec2}} recalls the dispersion relations for supersonically modulated media and presents a mode-coupling theory based on mode redistribution and degeneracy at spatial interfaces. In \textcolor{blue} {Section \ref{sec3}}, we quantitatively characterize the scattering behavior of bounded modulated media using the proposed theoretical framework, revealing a parametric amplification mechanism through Floquet-Bloch mode analysis. \textcolor{blue}{Section \ref{sec4}} presents a parametric study to assess the influence of modulation amplitude and modulation length on amplification performance. \textcolor{blue}{Section \ref{sec5}} validates the theoretical predictions via numerical simulations. Conclusions and future directions are discussed in \textcolor{blue}{Section \ref{sec6}}.

\section{Theoretical model}
\label{sec2}
\subsection{Problem description}
\label{subsec2.1}
\begin{figure}[h]   \centering
\includegraphics[width=1\linewidth]{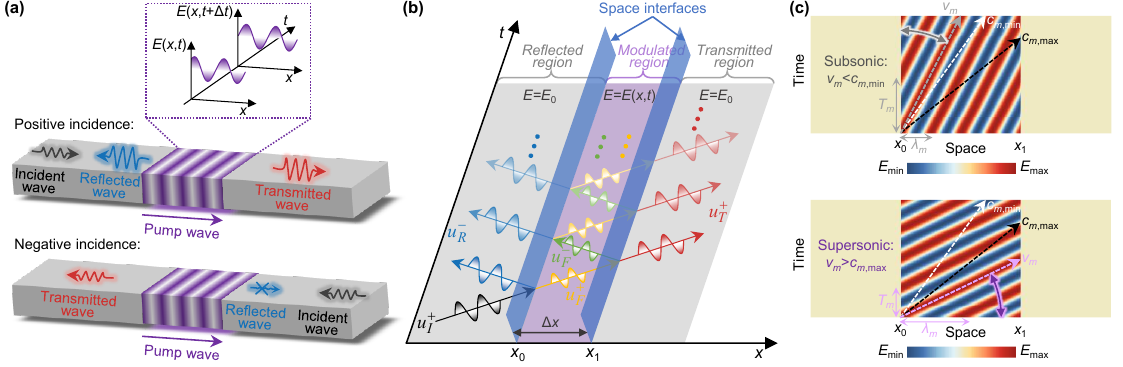}
\caption{ Schematic of longitudinal wave propagation in a slender rod embedded with a spatiotemporally modulated region. (a) Illustration of nonreciprocal parametric amplification. The inset depicts the elastic modulus \( E(x,t) \) modulated in both space and time in a traveling-wave form. (b) Conceptual sketch of spatial scattering of elastic longitudinal waves across a infinite spatiotemporally modulated region bounded by two space interfaces. The modulation length is defined as \( \Delta x = x_1-x_0 \). (c) Elastic modulus contours for subsonic and supersonic modulation regime. Black and white dashed lines represent the longitudinal wave propagation with maximum and minimum wave velocities, respectively. Gray and purple dashed lines represent the subsonic and supersonic modulation wave propagation, respectively.}
\label{Fig.1}
\end{figure}

We consider the propagation of longitudinal waves along an infinite homogeneous elastic rod featuring a finite-size modulated region, as illustrated in Fig.\ref{Fig.1}(a).  In the modulated region, the spatial variation of the material properties is synchronized with a temporal variation in the form of a pump wave (see the inset in \textcolor{blue}{Fig. \ref{Fig.1}}(a)). We consider two distinct wave incidence scenarios, referred to as positive and negative incidence, corresponding to cases where the incident wave propagates in the same or opposite direction as the pump wave, respectively (see \textcolor{blue}{Fig. \ref{Fig.1}}(a)).

As illustrated in \textcolor{blue}{Fig.~\ref{Fig.1}}(b), when an incident wave (denoted by \( u_I^+ \)) reaches the first spatial interface at \( x = x_0 \), which marks the beginning of the spatiotemporally modulated region, part of the wave is reflected, while the remainder propagates into the modulated region. When the forward-propagating wave reaches the second spatial interface at \( x = x_1 \), which marks the end of the modulated region, part of the wave is transmitted into the homogeneous medium, and part is reflected back into the modulated region. These forward- and backward-propagating waves undergo multiple scattering events at the two spatial interfaces, generating multiple transmitted (\( u_T^+ \)) and reflected (\( u_R^- \)) waves in the homogeneous regions.

Without loss of generality, the material modulation is implemented by prescribing a space-time-varying stiffness within the spatial region \( [x_0, x_1] \). The coordinates \( x_0 \) and \( x_1 \) mark the onset and termination of the spatiotemporal modulation. In what follows, we adopt a cosine-type modulation law, such that the elastic modulus over the entire spatial domain is given by:
\begin{equation}\label{Eq.1}
E(x,t)=
\begin{cases}
E_{0}, & x < x_{0}, \\
E_{0}\left[1 + \alpha_{m} \cos(\omega_{m}t - \kappa_{m}x)\right], & x_{0} \leq x \leq x_{1}, \\
E_{0}, & x > x_{1},
\end{cases}
\end{equation}
where \( E_{0} \) is the elastic modulus of the homogeneous medium, \( \alpha_{m} \) is the normalized modulation amplitude, and \( \omega_{m} \) and \( \kappa_{m} \) are the angular frequency and wavenumber of the modulation, respectively. The temporal and spatial periods of modulation are given by \( T_{m} = 2\pi/\omega_m \) and \( \lambda_{m} = 2\pi/\kappa_m \), respectively. The modulation velocity is defined as \( v_m = \omega_m/\kappa_m \). As such, the longitudinal wave velocity \( c_m(x,t) \) of modulated media is no longer a constant, but depends on both space and time, and varies within a range:
 \begin{equation}\label{Eq.2}
c_{m,\min} \leq c_m(x,t)=\sqrt{\frac{[1+\alpha_m\cos(\omega_mt-\kappa_mx)]E_0}{\rho_0}} \leq c_{m,\max},
\end{equation}
where \( c_{m,\min}=\sqrt{1-|\alpha_{m}|}c_{0} \) and \( c_{m,\max}=\sqrt{1+|\alpha_{m}|}c_{0} \) depend only on the modulation amplitude \( \alpha_m \) and define the minimum and maximum wave velocities in the modulated region, respectively. We display elastic modulus contours in a space-time domain for two representative modulation regimes, namely, subsonic modulation (\( v_m < c_{m,\min} \)) and supersonic modulation (\( v_m > c_{m,\max} \)), as depicted in \textcolor{blue}{Fig. \ref{Fig.1}}(c). Remarkably, the majority of studies on the scattering behavior of spatiotemporally modulated media have focused on the subsonic modulation regime, as it ensures a stable interaction between the guided wave and the modulation (or pump) wave \citep{cassedy1963dispersion,cassedy1967dispersion}.

In the following analysis, we focus on the supersonic modulation regime considering a sufficiently high modulation velocity, e.g.,  \( v_m > c_{m,\max} \).

\subsection{Wavenumber bandgaps in supersonic spatiotemporally modulated media}
\label{subsec2.2}

We begin by revisiting the dispersion characteristics of longitudinal waves propagating in a medium with supersonic spatiotemporal modulation. The governing equation for a slender elastic rod with a modulated elastic modulus \( E(x,t) \) is given by:
\begin{equation}\label{Eq.3}
\frac{\partial}{\partial x}\left[E(x,t)\frac{\partial u(x,t)}{\partial x}\right]
- \frac{\partial}{\partial t}\left[\rho_0 \frac{\partial u(x,t)}{\partial t}\right] = 0,
\end{equation}
where \( u(x,t) \) denotes the longitudinal displacement field along the \( x \)-direction, \( E(x,t) \) is the spatiotemporally varying elastic modulus, and \( \rho_0 \) is the constant mass density of the rod.

In the modulated region (\( x_0 \leq x \leq x_1 \)), due to the periodic nature in both space \( x \) and time \( t \) dimensions, \( E(x,t) \) can be written as a Fourier series:
\begin{equation}\label{Eq.4}
E(x,t)=\sum_{p=-P}^{+P}\hat{E}_{p}e^{ip(\omega_m t-\kappa_m x)},
\end{equation}
where \( \hat{E}_{p} \) is the Fourier coefficient, \( P \) is the truncation order of Fourier expansion. Since a cosine modulation function is chosen, \( \hat{E}_{p} = 0 \) for \( p = \pm 2, \pm 3, \cdots \). Hence, we truncate the order of Fourier expansion as \( p = 0, \pm 1 \), i.e., \( P = 1 \).

For homogeneous media, the dispersion relation can be simply obtained as \( \omega = \pm c_0\kappa \) from a typical governing equation of longitudinal waves with homogeneous media. Here, \( c_0 = \sqrt{E_0/\rho_{0}} \) is the longitudinal wave velocity in homogeneous media, and the "\( + \)" and "\( - \)" represent forward and backward propagating waves, respectively. For spatiotemporally modulated media, we employ a plane-wave expansion method (PWE) to derive the dispersion relation. The solution to Eq. \textcolor{blue}{(\ref{Eq.3})} is assumed in the generalized Floquet form and expressed as:
\begin{equation}\label{Eq.5}
u(x,t)=e^{i(\omega t-\kappa x)}\sum_{n=-\infty}^{+\infty}\hat{u}_{n}e^{in(\omega_m t-\kappa_m x)},
\end{equation}
where \( \hat{u}_{n} \) is the amplitude of the \( n^{\mathrm{th}} \)-order Floquet-Bloch mode. Substituting Eq. \textcolor{blue}{(\ref{Eq.5})} into Eq. \textcolor{blue}{(\ref{Eq.3})}, we obtain a quadratic eigenvalue problem (QEP). Considering the frequency-preserving energy transfer at space interfaces, we solve this QEP in terms of wavenumber \( \kappa \) for given frequency \( \omega \). Consequently, we obtain \( 4N+2 \) eigenvalues \( \hat{\kappa}_{s}^{+/-} \) and \( 4N+2 \) eigenvectors \( \mathbf{\hat{U}^{+/-}_s}=\left[\begin{matrix}{\hat{u}^{+/-}_{s,-N},}&{\cdots,}&{\hat{u}^{+/-}_{s,N}}\end{matrix}\right]^{T} \) (\( s=-N, \cdots, +N \)) for each given \( \omega \), by truncating the order of PWE to \( n =-N, \cdots, +N \). Here, the superscript "\( ^{+/-} \)" refers to the forward/backward Floquet-Bloch waves.

To generalize our analysis, we define the dimensionless modulation velocity as \( V = v_m / c_0 \), and introduce the dimensionless wavenumber \( \mu \) and frequency \( \Omega \) as follows:
\begin{equation}\label{Eq.6}
\mu = \frac{\kappa}{\kappa_{m}}, \quad \Omega = \frac{\omega}{c_{0} \kappa_{m}}.
\end{equation}

\begin{figure}[h!]   \centering
\includegraphics[width=1\linewidth]{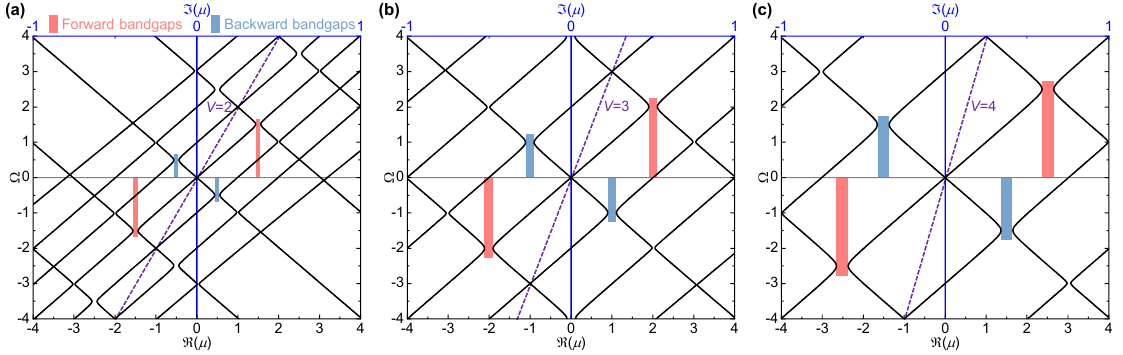}
\caption{Dispersion relations of elastic longitudinal waves under supersonic modulation for the same modulation amplitude \( \alpha_m = 0.3 \) but three different dimensionless modulation velocities: (a) \( V = 2 \), (b) \( V = 3 \), and (c) \( V = 4 \), respectively. Pink and blue rectangular shadings represent the first forward and backward wavenumber bandgaps of the fundamental branch, respectively.}
\label{Fig.2}
\end{figure}

To compute the dispersive properties, a truncation order of \( N = 3 \) is used, and the modulation amplitude is set to \( \alpha_m = 0.3 \). For this amplitude, the supersonic modulation regime corresponds to \( V > 1.14 \).
To account for all possible modes, we plot the dispersion relations in the complex \( \mu \) and real \( \Omega \) plane using a four-quadrant representation. \textcolor{blue}{Figs. \ref{Fig.2}}(a)--(c) show the analytical dispersion relations of the elastic longitudinal wave under supersonic modulation for three different dimensionless modulation velocities: \( V = 2 \), 3, and 4, respectively. These dispersion diagrams contain multiple branches distributed along an inclined direction with a slope of \( V \). Under supersonic modulation, adjacent wavenumber bands separate, forming wavenumber bandgaps (i.e. \( \mu \)-bandgaps), as highlighted by the pink and blue rectangular shadings in \textcolor{blue}{Figs. \ref{Fig.2}}(a)--(c). A comparison of the dispersion diagrams for different dimensionless modulation velocities \( V \) reveals that the \( \mu \)-bandgaps widen progressively with increasing \( V \).
From simple geometric relations, we determine that the first forward and backward \( \mu \)-bandgaps on the fundamental branch are located at \( (\pm\frac{V+1}{2},\pm\frac{V+1}{2}) \) and \( (\pm\frac{V-1}{2},\mp\frac{V-1}{2}) \), which depend only on the dimensionless velocity \( V \). Therefore, directional \( \mu \)-bandgaps can be customized for any arbitrary frequency by adjusting the dimensionless modulation velocity \( V \).

We remark that within the bandgaps, purely imaginary solutions for \( \mu \) are always absent, as shown by the blue lines in \textcolor{blue}{Figs. \ref{Fig.2}}(a)--(c). This suggests that, under supersonic modulation, spatial interfaces in spatiotemporally modulated media do not give rise to space-growing waves (i.e., forward evanescent waves), which are typically associated with a positive imaginary component of the wavenumber. As a result, the expected wave amplification phenomenon associated with supersonic modulation should be attributed to a different wave mechanism. In what follows, we describe this mechanism using mode-coupling theory and Floquet-Bloch mode analysis.

\subsection{Modeling of spatial scattering via mode-coupling theory}
\label{subsec2.3}
In this subsection, we adopt the theoretical framework introduced by \citet{yi2018reflection} to model the scattering behavior of elastic longitudinal waves propagating through a spatiotemporally modulated region under supersonic modulation. Without loss of generality, we illustrate the procedure for a dimensionless modulation velocity \( V = 2 \) and a normalized modulation amplitude \( \alpha_m = 0.3 \), which correspond to the dispersion curves presented in Fig.~\ref{Fig.2}(a).

\begin{figure}[h]   \centering
\includegraphics[width=1\linewidth]{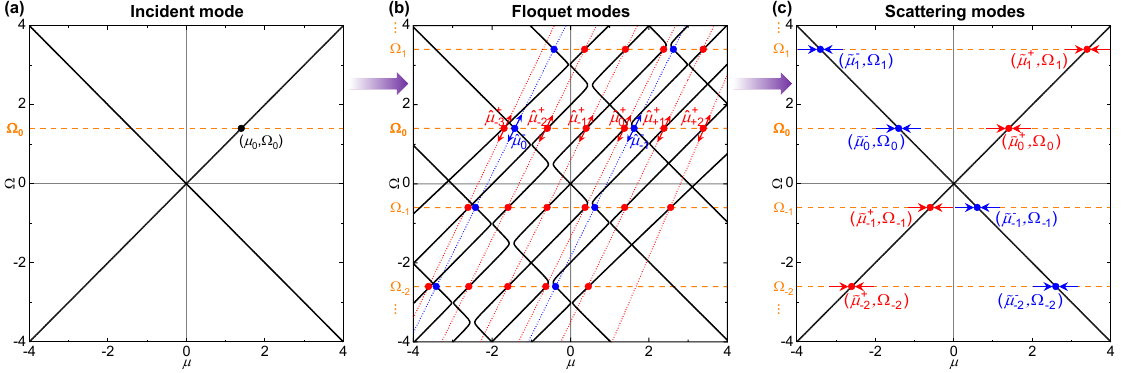}
\caption{Schematic diagram of the mode redistribution across space interfaces in spatiotemporally modulated media. (a) The incident mode is shown as a black dot on the dispersion diagram of a homogeneous medium. (b) The forward and backward Floquet-Bloch modes are shown as the red and blue dots on the dispersion diagram of a spatiotemporally modulated medium, respectively, where \( \Omega_n=\Omega_0+n \). The modulation parameters are selected as \( \alpha_m = 0.3 \), \( V = 2 \). The red and blue arrows are representative of the expansion of basic modes into forward and backward Floquet-Bloch modes. (c) The transmission and reflection modes are shown as the red and blue dots on the dispersion diagram of a homogeneous medium, respectively. The red and blue arrows are representative of the degeneration of Floquet-Bloch modes into transmission and reflection modes.}
\label{Fig.3}
\end{figure}

For simplicity, we consider a positive incident wave originating from the semi-infinite homogeneous medium on the left, described by a wavenumber-frequency pair \( (\kappa_0, \omega_0) \) (with the corresponding dimensionless form \( (\mu_0, \Omega_0) \)), as shown in \textcolor{blue}{Fig.~\ref{Fig.3}}(a). The incident wave can thus be expressed as
\begin{equation}\label{Eq.7}
u_{I}^{+} = A_0 e^{i(\omega_0 t - \kappa_0 x)},
\end{equation}
where \( A_0 \) is the wave amplitude, which is set to unity (\( A_0 = 1 \)) for computational convenience.

When the wave enters the modulated region (\( x_0 \leq x \leq x_1 \)), its temporal properties, such as the frequency, remain unchanged, while its spatial properties, such as the wavenumber, are altered. As a result, at the spatial interface (\( x = x_0 \)), a set of wave modes is generated with the same frequency \( \omega_0 \) (or \( \Omega_0 \) in dimensionless form), but with modified wavenumbers \( \kappa_0 \rightarrow \hat{\kappa}_s^{\pm} \) (or \( \mu_0 \rightarrow \hat{\mu}_s^{\pm} \)), as illustrated in \textcolor{blue}{Fig.~\ref{Fig.3}}(b). Based on the sign of their group velocities, these modes are classified as forward modes \( \hat{\kappa}_s^{+} \) (\( \hat{\mu}_s^{+} \)) and backward modes \( \hat{\kappa}_s^{-} \) (\( \hat{\mu}_s^{-} \)).

Due to the pumping effect arising from the spatiotemporal modulation within the modulated region, these basic modes are further expanded into a complete set of Floquet-Bloch modes, with wavenumbers shifted by \( n\kappa_m \) and frequencies shifted by \( n\omega_m \). Accordingly, we describe these forward and backward Floquet-Bloch modes by their wavenumber-frequency pairs \( (\hat{\kappa}_s^{\pm} + n\kappa_m,\, \omega_0 + n\omega_m) \), or in dimensionless form \( (\hat{\mu}_s^{\pm} + n,\, \Omega_0 + nV) \), where \( s, n = -N, \ldots, +N \). These modes are represented on the dispersion diagram of the spatiotemporally modulated medium as red and blue dots, respectively, in \textcolor{blue}{Fig.~\ref{Fig.3}}(b).

Hence, as a result of both the frequency-preserving energy transfer at spatial interfaces and the pumping effect of spatiotemporal modulation, the incident energy is redistributed into a series of Floquet-Bloch modes. The forward and backward Floquet-Bloch waves take the form:
\begin{equation}\label{Eq.8}
u_{F}^{+} = \sum_{s=-N}^{+N} \sum_{n=-N}^{+N} u_{F_{s,n}}^{+} = \sum_{s=-N}^{+N} B_{s} \sum_{n=-N}^{+N} \hat{u}_{s,n}^{+} e^{i[(\omega_{0}+n\omega_{m})t - (\hat{\kappa}_{s}^{+}+n\kappa_{m})x]},
\end{equation}
\begin{equation}\label{Eq.9}
u_{F}^{-} = \sum_{s=-N}^{+N} \sum_{n=-N}^{+N} u_{F_{s,n}}^{-} = \sum_{s=-N}^{+N} C_{s} \sum_{n=-N}^{+N} \hat{u}_{s,n}^{-} e^{i[(\omega_{0}+n\omega_{m})t - (\hat{\kappa}_{s}^{-}+n\kappa_{m})x]},
\end{equation}
where \( u_{F_{s,n}}^{\pm} \) represents the \( n^{\mathrm{th}} \)-order forward/backward Floquet-Bloch mode expanded from the \( s^{\mathrm{th}} \)-order basic mode. \( B_s \) and \( C_s \) are the amplitude coefficients of the forward and backward Floquet-Bloch modes, respectively. Here, \( s = -N, \ldots, +N \) denotes the order of the basic modes, and \( n = -N, \ldots, +N \) denotes the order of the Floquet-Bloch harmonics.

Note that in the supersonic modulation regime, no imaginary frequencies or wavenumbers are introduced; as a result, both forward and backward Floquet-Bloch modes always remain purely propagative rather than evanescent or exponentially growing. This behavior contrasts with recent observations in space-time modulated media bounded by temporal interfaces, where exponentially growing modes can arise \citep{ye2025nonreciprocal}.

As such, a distinct wave behavior governs the amplification mechanism. All Floquet-Bloch waves undergo repeated reflections at the two spatial interfaces located at \( x = x_0 \) and \( x = x_1 \), with only a fraction of their energy propagated into the adjacent semi-infinite homogeneous media on either side.

Within these adjacent media, the Floquet-Bloch modes degenerate into reflection and transmission modes, at \( x \leq x_0 \) and \( x \geq x_1 \), respectively.  In particular, all \( n^{\mathrm{th}} \)-order Floquet-Bloch harmonics, expanded from different basic modes, degenerate into \( n^{\mathrm{th}} \)-order scattering modes with preserved frequencies \( \omega_0 + n\omega_m \) (or \( \Omega_0 + nV \)) and modulated wavenumbers \( \hat{\kappa}_s^{\pm} + n\kappa_m \rightarrow \tilde{\kappa}_n^{\pm} \) (or \( \hat{\mu}_s^{\pm} + n \rightarrow \tilde{\mu}_n^{\pm} \)). We represent these transmission and reflection modes by their wavenumber-frequency pairs \( (\tilde{\kappa}_n^{\pm},\, \omega_0 + n\omega_m) \), or in dimensionless form \( (\tilde{\mu}_n^{\pm},\, \Omega_0 + nV) \), for \( n = -N, \ldots, +N \), as illustrated by the red and blue dots in \textcolor{blue}{Fig.~\ref{Fig.3}}(c). The transmitted and reflected waves are expressed as a sum of multiple scattered orders:
\begin{equation}\label{Eq.10}
u_{T}^{+} = \sum_{n=-N}^{+N} u_{T_{n}}^{+} = \sum_{n=-N}^{+N} \tilde{T}_{n} \, e^{i[(\omega_0 + n\omega_m)t - \tilde{\kappa}_{n}^{+} x]},
\end{equation}
\begin{equation}\label{Eq.11}
u_{R}^{-} = \sum_{n=-N}^{+N} u_{R_{n}}^{-} = \sum_{n=-N}^{+N} \tilde{R}_{n} \, e^{i[(\omega_0 + n\omega_m)t - \tilde{\kappa}_{n}^{-} x]},
\end{equation}
where \( u_{T_{n}}^{+} \) and \( u_{R_{n}}^{-} \) denote the \( n^\text{th} \)-order transmitted and reflected wave components, respectively; \( \tilde{T}_n \) and \( \tilde{R}_n \) are their corresponding amplitudes, and the wavenumbers are given by \( \tilde{\kappa}_{n}^{\pm} = \pm (\omega_0 + n\omega_m)/c_0 \), with \( c_0 \) being the wave speed in the homogeneous medium.

To solve the scattering problem, we impose displacement and stress continuity conditions at the space interfaces. For the space interface at \( x = x_0 \), the continuity conditions can be expressed as:
\begin{equation}\label{Eq.12}
{\left. u_{I}^{+} \right|}_{x={{x}_{0}}}+{\left. u_{R}^{-} \right|}_{x={{x}_{0}}}={\left. u_{F}^{+} \right|}_{x={{x}_{0}}}+{\left. u_{F}^{-} \right|}_{x={{x}_{0}}},
\end{equation}
\begin{equation}\label{Eq.13}
{{E}_{0}}{{\left. \frac{\partial u_{I}^{+}}{\partial x} \right|}_{x={{x}_{0}}}}+{{E}_{0}}{{\left. \frac{\partial u_{R}^{-}}{\partial x} \right|}_{x={{x}_{0}}}}=E({{x}_{0}},t){{\left. \frac{\partial u_{F}^{+}}{\partial x} \right|}_{x={{x}_{0}}}}+E({{x}_{0}},t){{\left. \frac{\partial u_{F}^{-}}{\partial x} \right|}_{x={{x}_{0}}}}. 
\end{equation}
Similarly, the continuity conditions for the space interface at \( x = x_1 \) are expressed as:
\begin{equation}\label{Eq.14}
{\left. u_{F}^{+} \right|}_{x={{x}_{1}}}+{\left. u_{F}^{-} \right|}_{x={{x}_{1}}}={\left. u_{T}^{+} \right|}_{x={{x}_{1}}}, \end{equation}
\begin{equation}\label{Eq.15}
E({{x}_{1}},t){{\left. \frac{\partial u_{F}^{+}}{\partial x} \right|}_{x={{x}_{1}}}}+E({{x}_{1}},t){{\left. \frac{\partial u_{F}^{-}}{\partial x} \right|}_{x={{x}_{1}}}}={{E}_{0}}{{\left. \frac{\partial u_{T}^{+}}{\partial x} \right|}_{x={{x}_{1}}}}.
\end{equation}

With some algebra, the scattering relation can be obtained and organized in matrix form as:
\begin{equation}\label{Eq.16}
\left( {{\mathbf{M}}_{\mathbf{1}}}{{A}_{0}}+{{\mathbf{M}}_{\mathbf{2}}}\mathbf{\tilde{R}} \right)-{{\mathbf{M}}_{\mathbf{3}}}\mathbf{M}_{\mathbf{4}}^{\mathbf{-1}}{{\mathbf{M}}_{\mathbf{5}}}\mathbf{\tilde{T}}=\mathbf{0},
\end{equation}
where, \( {\mathbf{M}}_{\mathbf{1}} \), \( {\mathbf{M}}_{\mathbf{2}} \), \( {\mathbf{M}}_{\mathbf{3}} \), \( {\mathbf{M}}_{\mathbf{4}} \) and \( {\mathbf{M}}_{\mathbf{5}} \) are transfer matrices. More details of the derivation and explicit expressions for the transfer matrices are provided in \textcolor{blue}{Appendix \ref{App.A}}. From Eq. \textcolor{blue}{(\ref{Eq.16})}, we can solve for \( 4J+2 \) unknowns, i.e., \( \tilde{T}_{-J}, \cdots, \tilde{T}_{J}, \tilde{R}_{-J}, \cdots, \tilde{R}_{J} \). Here, \( J = N-P \) is the truncation order of mode-coupling theory. Consequently, \( n^\mathrm{th} \)-order (\( -J\leq n \leq +J \)) transmission and reflection coefficients are given by:
\begin{equation}\label{Eq.17}
T_{n}=\left|\frac{\tilde{T}_{n}}{A_{0}}\right|,\quad R_{n}=\left|\frac{\tilde{R}_{n}}{A_{0}}\right|.
\end{equation}

\section{Analytical prediction of scattering behavior due to supersonic modulation}
\label{sec3}

In this section, we analytically predict the scattering behavior of longitudinal waves propagating through a supersonically modulated region by employing the mode-coupling theory described in \textcolor{blue}{Section~\ref{subsec2.2}}. Based on the previous discussion in \textcolor{blue}{Section~\ref{subsec2.1}}, we select the modulation parameters as follows: dimensionless modulation velocity \( V = 2 \), modulation wavenumber \( \kappa_m = 10\;\mathrm{rad/m} \), modulation frequency \( \omega_m = 20\;\mathrm{rad/s} \), normalized modulation amplitude \( \alpha_m = 0.3 \), and modulation length \( \Delta x = 3\lambda_m = 0.6\pi \). The reader can refer to the dispersion diagram shown in \textcolor{blue}{Fig.~\ref{Fig.2}}(b) with forward and backward \( \mu \)-bandgaps centered around \( \Omega_0 = 1.5 \) and \( \Omega_0 = 0.5 \), respectively.

\subsection{Transmission and reflection coefficients}
\label{subsec3.1}
\begin{figure}[h]   \centering
\includegraphics[width=1\linewidth]{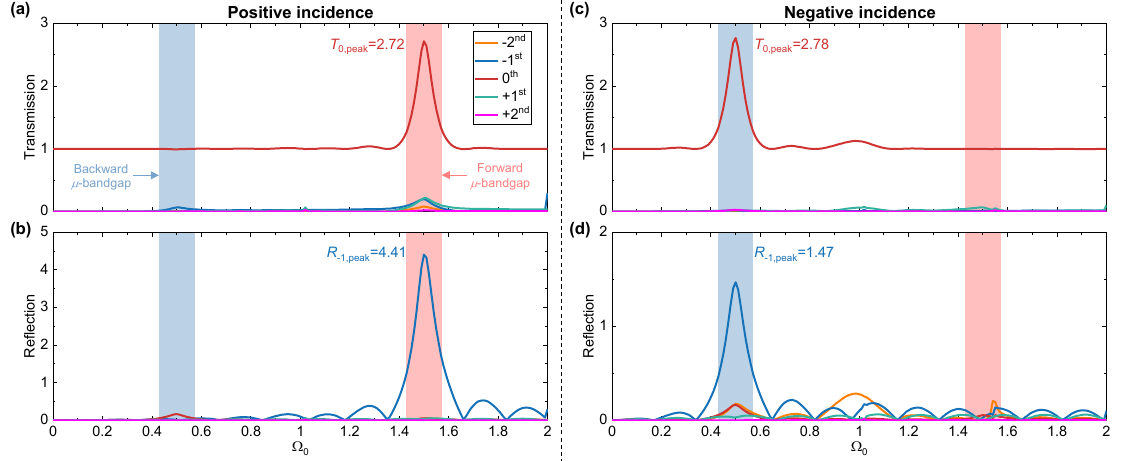}
\caption{ Analytical transmission and reflection coefficients. The analytical (a) transmission and (b) reflection coefficients for positive incidence. The analytical (c) transmission and (d) reflection coefficients for negative incidence.}
\label{Fig.4}
\end{figure}

The transmission and reflection coefficients computed using the mode-coupling theory for longitudinal waves incident from opposite directions are shown in \textcolor{blue}{Figs.~\ref{Fig.4}}(a)--(d). We begin by analyzing the case of positive incidence, with the corresponding transmission and reflection coefficients presented in \textcolor{blue}{Figs.~\ref{Fig.4}}(a) and (b), respectively.

Within the frequency range \( \Omega_0 \in [0,2] \), the transmission coefficient associated with the fundamental mode (i.e., \( 0^{\mathrm{th}} \)-order) remains equal to 1 at most frequencies, while all higher-order scattering coefficients are nearly zero. However, when the incident frequency falls within the region around \( \Omega_0 = 1.5 \), corresponding to the forward directional \( \mu \)-bandgap, both the \( 0^{\mathrm{th}} \)-order transmission and the \( -1^{\mathrm{st}} \)-order reflection coefficients increase sharply, reaching peak values of \( T_{0,\mathrm{peak}} = 2.72 \) and \( R_{-1,\mathrm{peak}} = 4.41 \), respectively. This behavior indicates that parametric amplification occurs through both transmission and reflection, accompanied by frequency conversion for the reflected mode, i.e., \( \tilde{\omega}_{-1} = \omega_0 - \omega_m \) (or \( \tilde{\Omega}_{-1} = \Omega_0 - V = -0.5 \)). Note that negative frequencies can always be mapped to their corresponding positive values by considering an opposite phase. Thus, the incident wave at \( \Omega_0 = 1.5 \) is reflected into the \( -1^{\mathrm{st}} \)-order mode with a down-converted absolute frequency of \( |\tilde{\Omega}_{-1}| = 0.5 \).

Let us now investigate the scattering properties for the negative incidence case. The corresponding transmission and reflection coefficients are shown in \textcolor{blue}{Figs.~\ref{Fig.4}}(c) and (d). For this case, we observe parametric amplification of the \( 0^{\mathrm{th}} \)-order transmission and the \( -1^{\mathrm{st}} \)-order reflection within the directional \( \mu \)-bandgap centered around \( \Omega_0 = 0.5 \). Both coefficients rise sharply, reaching peak values of \( T_{0,\mathrm{peak}} = 2.78 \) and \( R_{-1,\mathrm{peak}} = 1.47 \), respectively, at the center of the bandgap. Furthermore, the frequency of the \( -1^{\mathrm{st}} \)-order reflected wave is converted in the same manner as in the positive incidence case, i.e., \( \tilde{\omega}_{-1} = \omega_0 - \omega_m \) (or \( \tilde{\Omega}_{-1} = \Omega_0 - V = -1.5 \)). This results in a frequency up-conversion from the incident wave at \( \Omega_0 = 0.5 \) to the \( -1^{\mathrm{st}} \)-order reflected wave, which has an absolute frequency of \( |\tilde{\Omega}_{-1}| = 1.5 \).

Since the positions of the directional \( \mu \)-bandgaps depend solely on the dimensionless modulation velocity \( V \) (as shown in the analysis in \textcolor{blue}{Section~\ref{subsec2.1}}), parametric amplification and frequency conversion functionalities can be tuned for different excitation frequencies by adjusting \( V \). An extended investigation on the effect of the dimensionless modulation velocity \( V \) on the transmission and reflection coefficients is provided in \textcolor{blue}{Appendix~\ref{App.B}}, where it is shown how amplification can be induced at different incident frequencies by varying \( V \). In particular, the strongest parametric amplification within the considered range occurs at \( V = 2.45 \), yielding maximum peak values of \( T_{0,\mathrm{peak}} = 37.2 \) and \( R_{-1,\mathrm{peak}} = 58 \).

\subsection{Floquet-Bloch mode analysis}
\label{subsec3.2}
Following the approach of \cite{yi2018reflection} developed for subsonic modulated media, here we quantitatively investigate the Floquet-Bloch waves excited within the modulated medium to shed light on the mechanisms of parametric amplification and frequency conversion induced by supersonic modulation. Without loss of generality, we focus on the case of positive incidence and consider two representative incident frequencies, \( \Omega_0 = 1.0 \) and \( \Omega_0 = 1.5 \), which are located outside and at the center of the \( \mu \)-bandgap, respectively.

First, we extract the amplitude coefficients of the forward and backward Floquet-Bloch modes, denoted by \( F_{s,n}^{+} \) and \( F_{s,n}^{-} \), respectively. The amplitudes of the forward and backward Floquet-Bloch waves are given by \( B_s\hat{u}_{s,n}^{+} \) and \( C_s\hat{u}_{s,n}^{-} \), respectively. Thus, \( F_{s,n}^{+} \) and \( F_{s,n}^{-} \) can be expressed as:
\begin{equation}\label{Eq.18}
F_{s,n}^{+}=\left|\frac{B_{s}\hat{u}_{s,n}^{+}}{A_{0}}\right|,\quad F_{s,n}^{-}=\left|\frac{C_{s}\hat{u}_{s,n}^{-}}{A_{0}}\right|,
\end{equation}
where \( F_{s,n}^{+/-} \) corresponds to the amplitude coefficient of the \( n^{\mathrm{th}} \)-order forward/backward Floquet-Bloch mode expanded from the \( s^{\mathrm{th}} \)-order basic mode. Here, \( s,n = -J, \cdots, +J \), and \( B_s \) and \( C_s \) are obtained from Eqs. \textcolor{blue}{(\ref{Eq.A10})} and \textcolor{blue}{(\ref{Eq.A11})}.

\begin{figure}[h]   \centering
\includegraphics[width=1\linewidth]{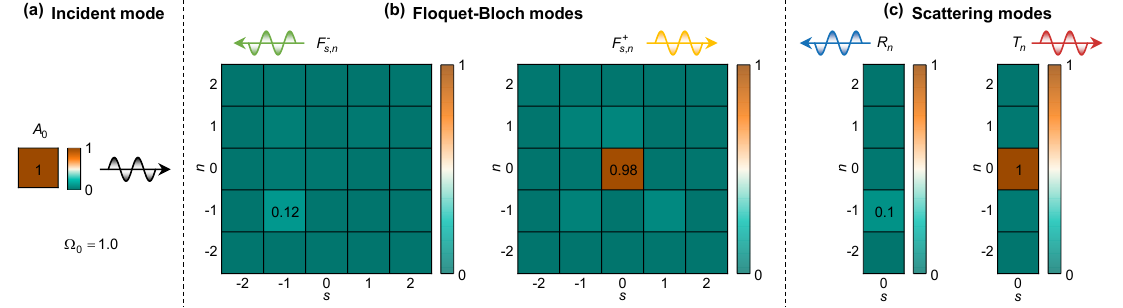}
\caption{ Wave modes excited in the waveguide for an incident frequency outside the \( \mu \)-bandgap, i.e., \( \Omega_0 = 1.0 \). The amplitude coefficients of the (a) incident mode, (b) Floquet-Bloch modes and (c) Scattering modes, respectively. The values in pixels indicate the relatively large amplitude coefficients in each mode groups.}
\label{Fig.5}
\end{figure}

We begin by considering an incident frequency outside the \( \mu \)-bandgap, i.e., \( \Omega_0 = 1.0 \). \textcolor{blue}{Figs.~\ref{Fig.5}}(a)--(c) show the amplitude coefficients of the incident mode, Floquet-Bloch modes, and scattering modes in the spatiotemporally modulated system. The color of each pixel represents the amplitude of a given wave mode relative to the incident mode. The relatively large magnitudes within each mode group are labeled with specific values.

In the modulated region, the incident energy is first decomposed into basic modes (\( u_{I}^{+} \to u_{F_{s,0}}^{\pm} \)), and simultaneously expanded into higher-order Floquet-Bloch modes (\( u_{F_{s,0}}^{\pm} \to u_{F_{s,n}}^{\pm} \)). In the case under analysis, nearly all incident energy is transferred to the fundamental Floquet-Bloch mode, \( u_{F_{0,0}}^{+} \), with an amplitude coefficient of \( F_{0,0}^{+} = 0.98 \), while only a small fraction is converted into backward Floquet-Bloch waves, with a maximum amplitude coefficient of \( F_{-1,-1}^{-} = 0.12 \), as shown in \textcolor{blue}{Fig.~\ref{Fig.5}}(b).

In the scattered fields, the energy of all \( n^{\mathrm{th}} \)-order Floquet-Bloch modes redistributes into \( n^{\mathrm{th}} \)-order scattering modes (\( u_{F_{s,n}}^{\pm} \to u_{T_n}^{+},\, u_{R_n}^{-} \)). As a result, the dominant Floquet-Bloch mode \( u_{F_{0,0}}^{+} \) is transmitted directly through the modulated region and converted into a forward scattering mode, specifically, the \( 0^{\mathrm{th}} \)-order transmission mode, without interference from other components, reaching a unitary transmission coefficient of \( T_0 = 1 \), as shown in \textcolor{blue}{Fig.~\ref{Fig.5}}(c). At the same time, reflection is negligible: the backward Floquet-Bloch modes carry minimal energy, resulting in a maximum amplitude coefficient of only \( R_{-1} = 0.1 \).

\begin{figure}[h]   \centering
\includegraphics[width=1\linewidth]{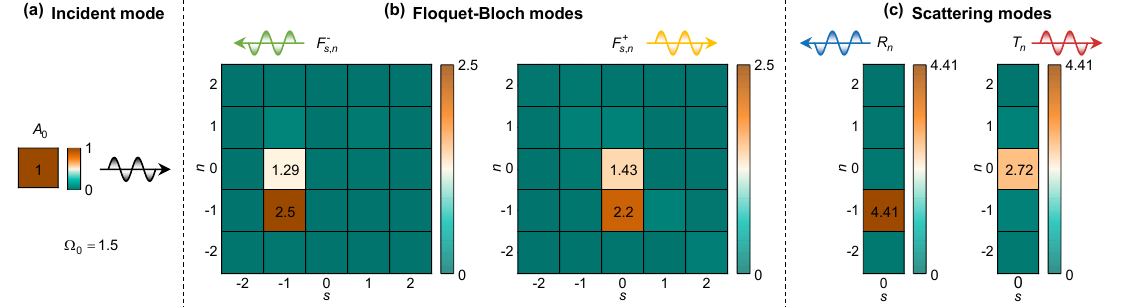}
\caption{Wave modes excited in the waveguide for an incident frequency within the \( \mu \)-bandgap, i.e., \( \Omega_0 = 1.5 \). The amplitude coefficients of the (a) incident mode, (b) Floquet-Bloch modes and (c) Scattering modes, respectively. The values in the pixels indicate the amplitude coefficients of dominant modes in different mode groups.}
\label{Fig.6}
\end{figure}

Using the same approach, we examine the mode redistribution associated with an incident wave whose frequency lies within the first forward \( \mu \)-bandgap, i.e., \( \Omega_0 = 1.5 \). The amplitude coefficients of the corresponding wave modes excited in the waveguide are shown in \textcolor{blue}{Figs.~\ref{Fig.6}}(a)--(c). Within the \( \mu \)-bandgap, certain forward and backward Floquet-Bloch modes are amplified by drawing energy from the supersonic modulation wave through the pumping effect.

In particular, the forward Floquet-Bloch wave field is dominated by the \( u_{F_{0,0}}^{+} \) and \( u_{F_{0,-1}}^{+} \) modes, with amplitude coefficients of \( F_{0,0}^{+} = 1.43 \) and \( F_{0,-1}^{+} = 2.2 \), respectively. Conversely, the backward Floquet-Bloch wave field is dominated by the \( u_{F_{-1,0}}^{-} \) and \( u_{F_{-1,-1}}^{-} \) modes, with amplitude coefficients of \( F_{-1,0}^{-} = 1.29 \) and \( F_{-1,-1}^{-} = 2.5 \), respectively, as shown in \textcolor{blue}{Fig.~\ref{Fig.6}}(b). These dominant Floquet-Bloch waves correspond precisely to the modes located at the left and right edges of the first forward and backward \( \mu \)-bandgaps.

This observation suggests that  supersonic modulation drives energy transfer between the forward and backward \( \mu \)-bandgaps in the direction of the modulation velocity \( v_m \). According to the previous analysis, all \( n^{\mathrm{th}} \)-order Floquet-Bloch modes interact within the modulated region and convert into \( n^{\mathrm{th}} \)-order scattering modes at the spatial interfaces while preserving their frequency. As a result, all Floquet-Bloch modes of order \( n = 0 \) are entirely transformed into the \( 0^{\mathrm{th}} \)-order forward scattering mode, while those of order \( n = -1 \) are exclusively transformed into the \( -1^{\mathrm{st}} \)-order backward scattering mode, i.e., \( u_{F_{0,0}}^{+},\, u_{F_{-1,0}}^{-} \to u_{T_0}^{+} \) and \( u_{F_{0,-1}}^{+},\, u_{F_{-1,-1}}^{-} \to u_{R_{-1}}^{-} \), as shown in \textcolor{blue}{Fig.~\ref{Fig.6}}(c).

Therefore, the transmitted field consists solely of the \( 0^{\mathrm{th}} \)-order transmitted wave with an amplified amplitude, while the reflected field contains only the \( -1^{\mathrm{st}} \)-order reflected wave, which exhibits both amplitude amplification and frequency conversion.

\begin{figure}[h]   \centering
\includegraphics[width=1\linewidth]{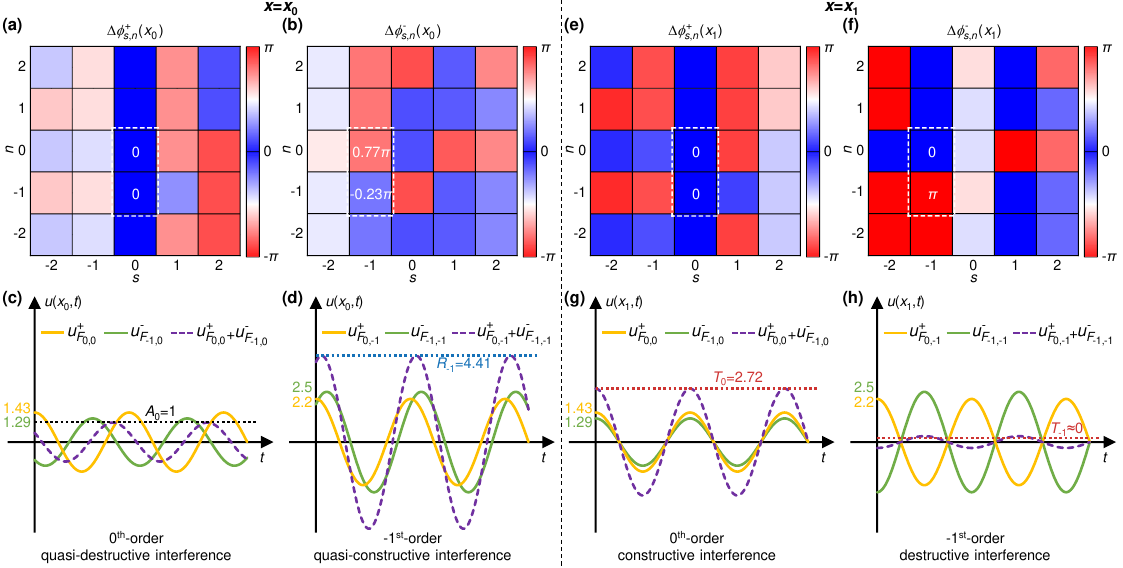}
\caption{Phase properties of Floquet-Bloch modes and interference behaviors between dominant counter-propagating Floquet-Bloch modes at \( x=x_0 \) and  \( x=x_1 \), respectively. Phase differences of (a) forward and (b) backward propagating Floquet-Bloch modes at \( x=x_0 \). (c) Quasi-destructive interference between dominant \( 0^\mathrm{th} \)-order counter-propagating Floquet-Bloch waves (i.e., \( u_{F_{0,0}}^{+} \) and \( u_{F_{-1,0}}^{-} \)) at \( x=x_0 \). (d) Quasi-constructive interference between dominant \( -1^\mathrm{st} \)-order counter-propagating Floquet-Bloch waves (i.e., \( u_{F_{0,-1}}^{+} \) and \( u_{F_{-1,-1}}^{-} \)) at \( x=x_0 \). Phase differences of (e) forward and (f) backward propagating Floquet-Bloch modes at \( x=x_1 \). (g) Constructive interference between dominant \( 0^\mathrm{th} \)-order counter-propagating Floquet-Bloch waves at \( x=x_1 \). (h) Destructive interference between dominant \( -1^\mathrm{st} \)-order counter-propagating Floquet-Bloch waves at \( x=x_1 \). The values in the pixels indicate the amplitude coefficients of dominant Floquet-Bloch modes. The orange and green solid lines represent the dominant \( 0^\mathrm{th} \)-order or \( -1^\mathrm{st} \)-order forward and backward propagating Floquet-Bloch waves, and the violet dashed line represents the corresponding interference wave.}
\label{Fig.7}
\end{figure}

To fully understand the interaction of these Floquet-Bloch modes within the \( \mu \)-bandgap, we  investigate their phase properties. We compute the phase difference between the \( n^{\mathrm{th}} \)-order Floquet-Bloch modes expanded from the \( s^{\mathrm{th}} \)-order basic mode and the \( 0^{\mathrm{th}} \)-order positive basic mode, namely:
\begin{equation}\label{Eq.19}
\Delta \phi_{s,n}^{+/-}=\phi_{s,n}^{+/-}-\phi_{0,n}^{+},
\end{equation}
where, \( \phi_{s,n}^{+}=\arctan \bigl[\Im(B_{s}\hat{u}_{s,n}^{+})/\Re({B_{s}\hat{u}_{s,n}^{+})}\bigr] \) and \( \phi_{s,n}^{-}=\arctan \bigl[\Im(C_{s}\hat{u}_{s,n}^{-})/\Re({C_{s}\hat{u}_{s,n}^{-})}\bigr] \) are the phases of forward Floquet-Bloch modes \( u_{F_{s,n}}^{+} \) and backward Floquet-Bloch modes \( u_{F_{s,n}}^{-} \), and \( \phi_{0,n}^{+} \) is the reference phase, i.e., \( \Delta \phi_{0,n}^{+}=0 \). Note that Floquet-Bloch modes with the same order \( n^\mathrm{th} \) have the same frequency \( \omega_0+n\omega_m \), thus \( \Delta \phi_{s,n}^{+/-} \) depends only on \( x \), i.e., \( \Delta \phi_{s,n}^{+/-}=\Delta \phi_{s,n}^{+/-}(x) \). The phase differences of all the Floquet-Bloch modes are extracted at both space interfaces at \( x=x_0 \) and \( x=x_1 \), and normalized to the range \( [-\pi,\pi] \), as shown in \textcolor{blue}{Figs. \ref{Fig.7}}(a), (b) and \textcolor{blue}{Figs. \ref{Fig.7}}(e), (f), respectively. For clarity, the phase differences of the dominant modes are highlighted using white dashed boxes (i.e., \( \Delta \phi_{0,0}^{+} \), \( \Delta \phi_{0,-1}^{+} \), \( \Delta \phi_{-1,0}^{-} \), and \( \Delta \phi_{-1,-1}^{-} \)).

Let us first examine the interaction of dominant Floquet-Bloch modes at the interface \( x = x_0 \). From Figs.~\ref{Fig.7}(a) and \ref{Fig.7}(b), the dominant \( 0^{\mathrm{th}} \)-order Floquet-Bloch modes, namely \( u_{F_{0,0}}^{+} \) and \( u_{F_{-1,0}}^{-} \), are nearly out of phase, exhibiting a phase difference of \( \Delta \phi_{-1,0}^{-}(x_0) = 0.77\pi \). As a result, quasi-destructive interference occurs between these modes, with the amplitude of the resulting superposed wave given by \( \bigl|u_{F_{0,0}}^{+}(x_0,t) + u_{F_{-1,0}}^{-}(x_0,t)\bigr| \approx 1 \), as shown in Fig.~\ref{Fig.7}(c). As expected, the amplitude of the superposition at \( x = x_0 \) approximately matches that of the incident wave.

Conversely, the dominant \( -1^{\mathrm{st}} \)-order Floquet-Bloch modes, i.e., \( u_{F_{0,-1}}^{+} \) and \( u_{F_{-1,-1}}^{-} \), are nearly in phase, with a phase difference of \( \Delta \phi_{-1,-1}^{-}(x_0) = -0.23\pi \). Consequently, quasi-constructive interference arises, resulting in an amplified superposition with an amplitude comparable to that of the \( -1^{\mathrm{st}} \)-order reflected mode, namely \( \bigl|u_{F_{0,-1}}^{+}(x_0,t) + u_{F_{-1,-1}}^{-}(x_0,t)\bigr| \approx R_{-1} \), as illustrated in Fig.~\ref{Fig.7}(d).

In the same way, we examine the interaction of dominant Floquet-Bloch modes at \( x=x_1 \). As shown in \textcolor{blue}{Figs. \ref{Fig.7}}(e) and (f), the dominant \( 0^{\mathrm{th}} \)-order Floquet-Bloch modes are completely in phase, with a phase difference of \( \Delta \phi_{-1,0}^{-}(x_1) = 0 \), whereas the dominant \( -1^{\mathrm{st}} \)-order Floquet-Bloch modes are fully out of phase, with \( \Delta \phi_{-1,-1}^{-}(x_1) = \pi \). Consequently, the dominant \( 0^{\mathrm{th}} \)-order Floquet-Bloch modes constructively interfere, producing a \( 0^{\mathrm{th}} \)-order transmitted mode with an amplified amplitude, \( \bigl|u_{F_{0,0}}^{+}(x_1,t)+u_{F_{-1,0}}^{-}(x_1,t)\bigr| \approx T_0 \), as shown in \textcolor{blue}{Fig. \ref{Fig.7}}(g). In contrast, the dominant \( -1^{\mathrm{st}} \)-order Floquet-Bloch modes destructively interfere, suppressing the \( -1^{\mathrm{st}} \)-order transmission, i.e., \( \bigl|u_{F_{0,-1}}^{+}(x_1,t)+u_{F_{-1,-1}}^{-}(x_1,t)\bigr| \approx 0 \), as shown in \textcolor{blue}{Fig. \ref{Fig.7}}(h). The above analysis shows that the amplification and frequency conversion induced by supersonic modulation arise from distinct interference mechanisms between forward and backward Floquet-Bloch modes at the two space interfaces. The synergy of constructive and destructive interference selectively enhances the \( 0^\mathrm{th} \)-order transmitted mode and the \( -1^\mathrm{st} \)-order reflected mode, while simultaneously suppressing the \( -1^\mathrm{st} \)-order transmitted mode and the \( 0^\mathrm{th} \)-order reflected mode.

\section{Parametric study}
\label{sec4}
We investigate how modulation parameters influence amplification performance at a fixed operating frequency. In particular, we focus on the effects of modulation length \( \Delta x \) and modulation amplitude \( \alpha_m \) on the overall amplification behavior.

\begin{figure}[h]   \centering
\includegraphics[width=1\linewidth]{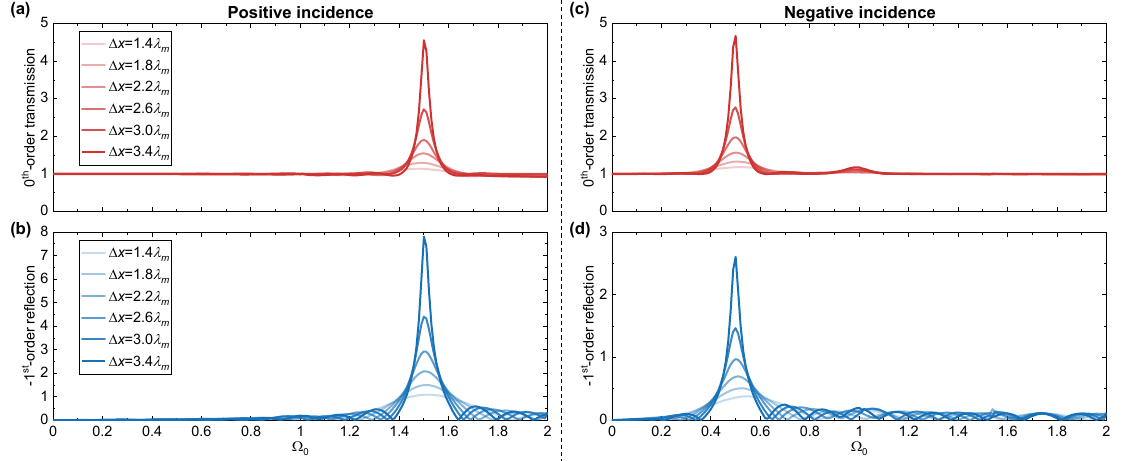}
\caption{Effect of modulation length on amplification performance. Other modulation parameters are fixed as: \( \kappa_m=10\;\mathrm{rad/m} \), \( \omega_m=20\;\mathrm{rad/s} \), \( \alpha_m=0.3 \). (a) The \( 0^\mathrm{th} \)-order transmission coefficients and (b) the \( -1^\mathrm{st} \)-order reflection coefficients for different modulation lengths under positive incidence. (c) The \( 0^\mathrm{th} \)-order transmission coefficients and (d) the \( -1^\mathrm{st} \)-order reflection coefficients for different modulation lengths under negative incidence.}
\label{Fig.8}
\end{figure}

First, we investigate the influence of modulation length while keeping the other modulation parameters fixed (\( \kappa_m = 10\;\mathrm{rad/m} \), \( \omega_m = 20\;\mathrm{rad/s} \), \( \alpha_m = 0.3 \)). Since the locations of the \( \mu \)-bandgaps are solely determined by the dimensionless modulation velocity, the first forward and backward \( \mu \)-bandgaps remain centered at \( \Omega_0 = 1.5 \) and \( \Omega_0 = 0.5 \), respectively. For simplicity, we focus on the dominant scattering modes, namely the \( 0^{\mathrm{th}} \)-order transmission and the \( -1^{\mathrm{st}} \)-order reflection. Their amplitudes under positive incidence, for various modulation lengths and excitation frequencies, are computed using mode-coupling theory and shown in \textcolor{blue}{Figs.~\ref{Fig.8}}(a) and (b). It is evident that both transmission and reflection exhibit amplification near \( \Omega_0 = 1.5 \), and that this amplification becomes more pronounced as the modulation length increases. The corresponding coefficient profiles for negative incidence are presented in \textcolor{blue}{Figs.~\ref{Fig.8}}(c) and (d), showing a similar dependence on modulation length, but centered within the backward \( \mu \)-bandgap at \( \Omega_0 = 0.5 \). Increasing the modulation length enables greater energy transfer from the modulated wave into the guided wave, promoting the excitation of higher-amplitude Floquet-Bloch modes. This, in turn, leads to stronger transmission and reflection due to constructive interference at the spatial interfaces.

\begin{figure}[h]   \centering
\includegraphics[width=1\linewidth]{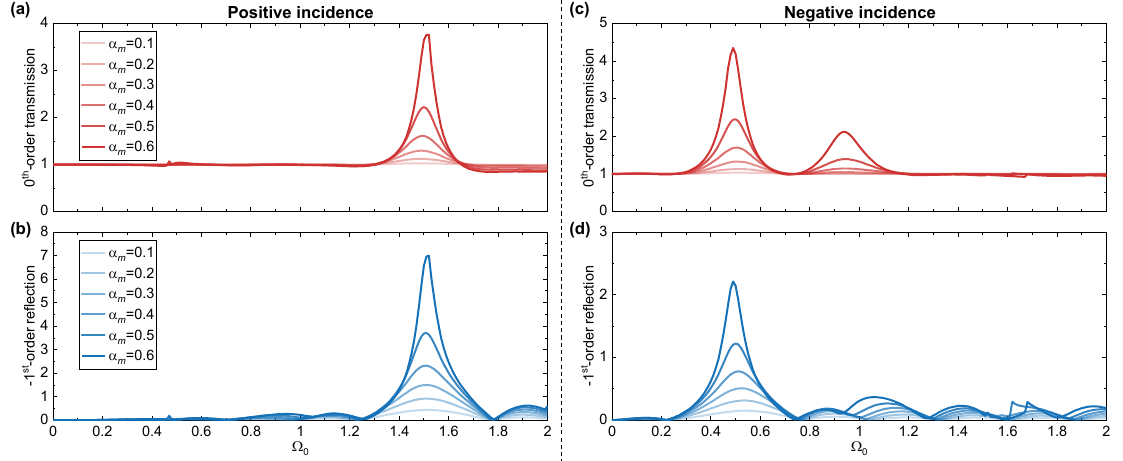}
\caption{Effect of modulation amplitude on amplification performance. Other modulation parameters are fixed as: \( \kappa_m=10\;\mathrm{rad/m} \), \( \omega_m=20\;\mathrm{rad/s} \), \( \Delta x=1.8\lambda_m \). (a) The \( 0^\mathrm{th} \)-order transmission coefficients and (b) the \( -1^\mathrm{st} \)-order reflection coefficients for different modulation amplitudes under positive incidence. (c) The \( 0^\mathrm{th} \)-order transmission coefficients and (d) the \( -1^\mathrm{st} \)-order reflection coefficients for different modulation amplitudes under negative incidence.}
 \label{Fig.9}
\end{figure}

Next, we consider the effect of modulation amplitude on the amplification, while keeping the other modulation parameters fixed as \( \kappa_m = 10\;\mathrm{rad/m} \), \( \omega_m = 20\;\mathrm{rad/s} \), and \( \Delta x = 1.8\lambda_m \). The \( 0^{\mathrm{th}} \)-order transmission and \( -1^{\mathrm{st}} \)-order reflection coefficients are analytically computed for various modulation amplitudes under both positive and negative incidence, as shown in \textcolor{blue}{Figs.~\ref{Fig.9}}(a)--(d). The transmission and reflection coefficients show amplification peaks at \( \Omega_0 = 1.5 \) and \( \Omega_0 = 0.5 \) for positive and negative incidence, respectively. These amplification peaks grow with increasing modulation amplitude due to enhanced energy pumping, which excites higher-amplitude Floquet-Bloch modes. In particular, for negative incidence, a secondary amplification peak also appears at \( \Omega_0 = 1.0 \) as the modulation amplitude increases. This behavior is attributed to the gradual opening of the second backward \( \mu \)-bandgap.

Overall, these results demonstrate that by appropriately tuning the modulation amplitude and modulation length, the parametric amplification performance can be effectively tailored to meet specific design objectives. Nonetheless, it is important to emphasize that modulation parameters must be carefully selected, as improper combinations, such as excessive modulation amplitude or length, can trigger parametric instability. This phenomenon is intrinsic to spatiotemporally modulated systems, particularly under supersonic modulation, as highlighted in recent studies on time-varying oscillators and discrete systems with spatiotemporal modulation \citep{jin2024exceptional,wu2025parametric}. 

When parametric instability arises, it excites parameter-resonant modes within the system, resulting in exponential growth of the response amplitude over time. Under such conditions, conventional scattering coefficients become invalid, as the system no longer reaches a steady state. To illustrate this behavior, Appendix~\ref{App.C} presents numerical wavefields that capture the system’s unstable response under higher modulation amplitudes and extended modulation lengths.

\section{Numerical verification}
\label{sec5}
In this section, we perform finite difference time domain (FDTD) simulations to numerically verify the spatial scattering behaviors predicted by the theoretical model. The modulation parameters are defined to be consistent with the settings in the analytical calculation. We consider and discretize a space domain of length \( X=50\pi=250\lambda_m \), and a time domain of length \( T=40\pi=400T_m \). The supersonic spatiotemporal modulation is activated at \( x_0=0.5X=125\lambda_m \), and deactivated at \( x_1=x_0+\Delta x \).

First, we perform simulations for the case of positive incidence. At \( t = 0 \), a tone burst excitation \( u(x) = 0.5\left(1 - \cos\frac{\kappa_0 x + \delta\pi}{\delta}\right) e^{i(\kappa_0 x + \delta\pi)} \), for \( 0 \leq x \leq \frac{\delta\pi}{\kappa_0} \), is imposed, where \( \delta \) controls the spatial span of the incident wave and the central frequency is given by \( \omega_0 = c_0 \kappa_0 \). Transient wavefields for different incident frequencies are obtained by injecting wave packets with varying values of \( \kappa_0 \). The corresponding scattering spectra are computed via Fast Fourier Transform (FFT).

\begin{figure}[h]   \centering
\includegraphics[width=1\linewidth]{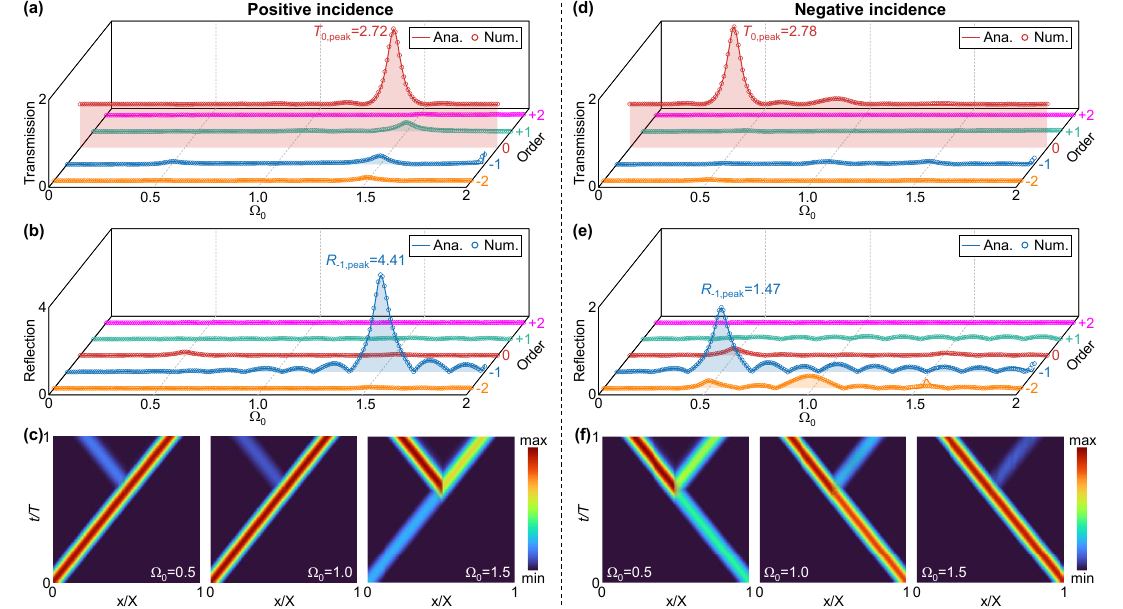}
\caption{Numerical verification for spatial scattering behaviors induced by supersonic spatiotemporal modulation. (a) Transmission and (b) reflection coefficients for positive incidence. (c) Time-history amplitude wavefields for positive incidence at different central frequencies \( \Omega_0 = 0.5 \), 1.0, and 1.5. (d) Transmission and (e) reflection coefficients for negative incidence. (f) Time-history amplitude wavefields for negative incidence at different central frequencies \( \Omega_0 = 0.5 \), 1.0, and 1.5.}
\label{Fig.10}
\end{figure}

The numerical transmission and reflection coefficients for different incident frequencies are shown in \textcolor{blue}{Figs.~\ref{Fig.10}}(a) and (b), respectively. For comparison, the analytical results obtained using mode-coupling theory are also provided. The numerical results agree well with the analytical predictions across all scattering orders. As expected, the \( 0^{\mathrm{th}} \)-order transmission and \( -1^{\mathrm{st}} \)-order reflection are amplified within the forward \( \mu \)-bandgap, reaching peak values at \( \Omega_0 = 1.5 \), consistent with the predicted coefficients \( T_{0,\mathrm{peak}} = 2.72 \) and \( R_{-1,\mathrm{peak}} = 4.41 \).

\textcolor{blue}{Figure~\ref{Fig.10}}(c) shows the time-history amplitude wavefields obtained from FDTD simulations for different central frequencies, namely \( \Omega_0 = 0.5 \), 1.0, and 1.5, revealing distinct scattering behaviors. It is clearly observed that at \( \Omega_0 = 0.5 \) and 1.0, the incident wave is transmitted without parametric amplification, and with negligible reflection. In contrast, at \( \Omega_0 = 1.5 \), both the transmitted and reflected waves experience parametric amplification, with amplitudes significantly exceeding that of the incident wave.

To investigate negative incidence, we impose an excitation from the opposite direction at \( t=0 \) using the displacement profile: \( u(x)=0.5\bigl[1-\cos\frac{\kappa_0(X-x)+\delta\pi}{\delta}\bigr]e^{i[\kappa_0(X-x)+\delta\pi]} \), for \( X-\frac{\delta\pi}{\kappa_0} \leq x \leq X \). The numerical transmission and reflection coefficients for negative incidence are compared with the analytical results in \textcolor{blue}{Figs. \ref{Fig.10}}(d) and (e), respectively. Again, the numerical results are consistent with the analytical results. As anticipated, the \( 0^\mathrm{th} \)-order transmission and \( -1^\mathrm{st} \)-order reflection show parametric amplification near \( \Omega_0 = 0.5 \), with predicted peak values of \( T_{0,peak} = 2.78 \) and \( R_{-1,peak} = 1.47 \). The corresponding time-history amplitude wavefields for different excitation frequencies centered at \( \Omega_0 = 0.5 \), 1.0, and 1.5 are shown in \textcolor{blue}{Fig. \ref{Fig.10}}(f). In this case, the incident wave centered at \( \Omega_0 = 0.5 \) is amplified through both transmission and reflection, while the incident waves centered at \( \Omega_0 = 1.0 \) and 1.5 are completely transmitted without amplification.

The comparison between positive and negative incidence cases reveals the nonreciprocal wave behavior within the directional \( \mu \)-bandgaps associated with supersonic modulation. Further evidence of non-reciprocal phenomena for broadband excitations is provided in \textcolor{blue}{Appendix~\ref{App.D}}. 

The influence of modulation parameters on amplification performance is further examined through numerical simulations, which confirm that increasing modulation length and modulation amplitude enhances amplification, as detailed in \textcolor{blue}{Appendix~\ref{App.E}}.

\section{Conclusions}
\label{sec6}
We have theoretically and numerically investigated an elastic waveguide containing a finite space-time modulated region, where the elastic properties vary in both space and time under supersonic modulation. We first computed the dispersion diagram in the presence of such modulation, identifying directional wavenumber bandgaps. Employing a theoretical framework based on mode-coupling theory, we analytically predicted the transmission and reflection coefficients of elastic longitudinal waves propagating through the spatially bounded, supersonically modulated medium. For excitations lying within the wavenumber bandgaps, we observed nonreciprocal parametric amplification in both transmission and reflection, accompanied by frequency conversion in the reflected waves. Analysis of the phase behavior of Floquet-Bloch modes confirmed that the observed amplification arises from (quasi-)constructive interference between counter-propagating Floquet-Bloch components at the spatial interfaces. Parametric studies further revealed that increasing the modulation length and amplitude enhances the amplification effect. These analytical results were validated through FDTD simulations, which confirmed the occurrence of nonreciprocal phenomena in supersonically modulated elastic systems.

Supersonically modulated continuous systems have received limited attention, largely due to the prevailing belief that they inherently lead to unstable, unbounded responses. However, our study demonstrates that by introducing spatial interfaces, these growing waves can be effectively confined, enabling stable parametric amplification. This offers a promising strategy for elastic wave control, enabling functionalities such as signal enhancement and frequency conversion. The generality of our theoretical framework suggests that the findings may be extended to other wave domains, including electromagnetic, acoustic, and various classes of elastic media, paving the way for future experimental implementations.

%% We hope that this initial exploration of supersonic modulation will motivate further investigations into this emerging wave control strategy and its potential for stable wave amplification. 
%Although parametric instability in spatiotemporally modulated discrete systems, both single- and multi-degree-of-freedom, has been explored in recent studies \citep{jin2024exceptional,wu2025parametric}, it remains an open challenge in the context of spatiotemporally modulated continuous systems. 
%Future efforts will focus on systematically exploring the parametric instability in such continuous systems with various spatial boundary conditions. This includes identifying critical thresholds, mapping instability regimes, and quantitatively assessing how modulation parameters influence the onset and characteristics of parametric instability.

\section*{CRediT authorship contribution statement
}
\textbf{Yingrui Ye:} Conceptualization, Investigation, Formal analysis and modeling, Methodology, Visualization, Writing - original draft, Funding acquisition.
\textbf{Chunxia Liu:} Conceptualization, Investigation, Formal analysis and modeling, Methodology.
%\textbf{Alessandro Marzani:} Conceptualization, Investigation, Formal analysis, Project administration, Supervision, Writing - review \& editing.
%\textbf{Emanuele Riva:} Formal analysis, Supervision, Writing - review \& editing.
\textbf{Xiaopeng Wang:} Conceptualization, Project administration, Supervision, Writing - review \& editing.
\textbf{Antonio Palermo:} Conceptualization, Investigation, Formal analysis, Project administration, Supervision, Writing - review \& editing, Funding acquisition.

\section*{Declaration of competing interest}
The authors declare that they have no known competing financial interests or personal relationships that could have appeared to influence the work reported in this paper.

\section*{Acknowledgments}
A.P. acknowledges the funding received from the Italian Ministry of University and Research (MUR) for the "EXTREME" project (grant agreement 2022EZT2ZE, CUP: J53C24002870006). Y.Y. gratefully acknowledges the support from the China Scholarship Council (CSC Grant No. 202306280145).

\appendix
\section{Detailed derivations and transfer matrices of mode-coupling theory for modeling spatial scattering behavior}
\label{App.A}
\renewcommand{\thefigure}{A.\arabic{figure}}
\setcounter{figure}{0}

For simplicity, we set \( x_0 \) as the origin of the spatial coordinate, indicating \( x_0 = 0 \) and \( x_1 = \Delta x \). Substituting Eqs. \textcolor{blue}{(\ref{Eq.7})}, \textcolor{blue}{(\ref{Eq.8})}, \textcolor{blue}{(\ref{Eq.9})}, \textcolor{blue}{(\ref{Eq.10})} and \textcolor{blue}{(\ref{Eq.11})} into Eqs. \textcolor{blue}{(\ref{Eq.12})} and \textcolor{blue}{(\ref{Eq.13})}, using the Fourier expansion of the elastic modulus shown in Eq. \textcolor{blue}{(\ref{Eq.4})}, we obtain:
\begin{small}
\begin{equation}\label{Eq.A1}\tag{A1}
{{A}_{0}}{{e}^{i{{\omega }_{0}}t}}+\sum\limits_{n=-N}^{+N}{{\tilde{R}_{n}}{{e}^{i{{\omega }_{n}}t}}}=\sum\limits_{s=-N}^{+N}{\left( {{B}_{s}}\sum\limits_{n=-N}^{+N}{\hat{u}_{s,n}^{+}{{e}^{i{{\omega }_{n}}t}}}+{{C}_{s}}\sum\limits_{n=-N}^{+N}{\hat{u}_{s,n}^{-}{{e}^{i{{\omega }_{n}}t}}} \right)},
\end{equation}
\end{small}
\begin{small}
\begin{equation}\label{Eq.A2}\tag{A2}
{{E}_{0}}\left( {{\kappa }_{0}}{{A}_{0}}{{e}^{i{{\omega }_{0}}t}}+\sum\limits_{n=-N}^{+N}{\tilde{\kappa }_{n}^{-}{\tilde{R}_{n}}{{e}^{i{{\omega }_{n}}t}}} \right)=\sum\limits_{s=-N}^{+N}{\left( {{B}_{s}}\sum\limits_{n=-N}^{+N}{\sum\limits_{p=-P}^{+P}{{{{\hat{E}}}_{p}}\hat{\kappa }_{s,n}^{+}\hat{u}_{s,n}^{+}{{e}^{i{{\omega }_{n+p}}t}}}}+{{C}_{s}}\sum\limits_{n=-N}^{+N}{\sum\limits_{p=-P}^{+P}{{{{\hat{E}}}_{p}}\hat{\kappa }_{s,n}^{-}\hat{u}_{s,n}^{-}{{e}^{i{{\omega }_{n+p}}t}}}} \right)}, 
\end{equation}
\end{small}
where, \( \omega_{n}=\omega_{0}+n\omega_{m} \), and \( \hat{\kappa}_{s,n}^{+/-}=\hat{\kappa}_{s}^{+/-}+n\kappa_{m} \).

Similarly, substituting Eqs. \textcolor{blue}{(\ref{Eq.7})}, \textcolor{blue}{(\ref{Eq.8})}, \textcolor{blue}{(\ref{Eq.9})}, \textcolor{blue}{(\ref{Eq.10})} and \textcolor{blue}{(\ref{Eq.11})} into Eqs. \textcolor{blue}{(\ref{Eq.14})} and \textcolor{blue}{(\ref{Eq.15})}, we get:
\begin{small}
\begin{equation}\label{Eq.A3}\tag{A3}
\sum\limits_{s=-N}^{+N}{\left( {{B}_{s}}\sum\limits_{n=-N}^{+N}{\hat{u}_{s,n}^{+}{{e}^{i({{\omega }_{n}}t-\hat{\kappa }_{s,n}^{+}\Delta x)}}}+{{C}_{s}}\sum\limits_{n=-N}^{+N}{\hat{u}_{s,n}^{-}{{e}^{i({{\omega }_{n}}t-\hat{\kappa }_{s,n}^{-}\Delta x)}}} \right)}=\sum\limits_{n=-N}^{+N}{{\tilde{T}_{n}}{{e}^{i({{\omega }_{n}}t-\tilde{\kappa }_{n}^{+}\Delta x)}}},
\end{equation}
\end{small}
\begin{small}
\begin{equation}\label{Eq.A4}\tag{A4}
\sum\limits_{s=-N}^{+N}{\left( {{B}_{s}}\sum\limits_{n=-N}^{+N}{\sum\limits_{p=-P}^{+P}{{{{\hat{E}}}_{p}}\hat{\kappa }_{s,n}^{+}\hat{u}_{s,n}^{+}{{e}^{i({{\omega }_{n+p}}t-\hat{\kappa }_{s,n+p}^{+}\Delta x)}}}}+{{C}_{s}}\sum\limits_{n=-N}^{+N}{\sum\limits_{p=-P}^{+P}{{{{\hat{E}}}_{p}}\hat{\kappa }_{s,n}^{-}\hat{u}_{s,n}^{-}{{e}^{i({{\omega }_{n+p}}t-\hat{\kappa }_{s,n+p}^{-}\Delta x)}}}} \right)}={{E}_{0}}\sum\limits_{n=-N}^{+N}{\tilde{\kappa }_{n}^{+}{\tilde{T}_{n}}{{e}^{i({{\omega }_{n}}t-\tilde{\kappa }_{n}^{+}\Delta x)}}}.
\end{equation}
\end{small}

We note that the temporal components of all the above modes are orthogonal, namely:
\begin{equation}\label{Eq.A5}\tag{A5}
\frac{1}{T_{m}}\int_{-T_{m}/2}^{T_{m}/2}\psi_{l}(t)\cdot\psi_{q}^{*}(t)dt=
\begin{cases}
1&l=q\\
0&l\neq q
\end{cases},
\end{equation}
where, \( \psi_{l}(t)=e^{-i(\omega_{0}+l\omega_{m})t} \), \( \psi_{q}(t)=e^{-i(\omega_{0}+q\omega_{m})t} \) and the superscript \( ^* \) represents complex conjugate.

By multiplying by \( e^{i\omega_{j}t} \) (\( j = -J, \cdots, +J \)), and utilizing the orthogonality relation shown in Eq. \textcolor{blue}{(\ref{Eq.A5})}, Eqs. \textcolor{blue}{(\ref{Eq.A1})}, \textcolor{blue}{(\ref{Eq.A2})}, \textcolor{blue}{(\ref{Eq.A3})} and \textcolor{blue}{(\ref{Eq.A4})} can each be decomposed into a system of \( 2J+1 \) equations. Applying this operation to Eq. \textcolor{blue}{(\ref{Eq.A2})} or \textcolor{blue}{(\ref{Eq.A4})} yields terms involving eigenvalues \( \hat{\kappa }_{s,j-p}^{+/-} \) and eigenvectors \( \hat{u}_{s,j-p}^{+/-} \). Given that the truncation order of PWE is \( n = -N, \cdots, +N \), the index \( j-p \) is constrained by \( -N \leq j-p \leq +N \). Recalling that the truncation order of the Fourier expansion of \( E(x,t) \) is \( p = -P, \cdots, +P \), the index \( j \) should satisfy \( -(N-P) \leq j \leq +(N-P) \), i.e., \( J \leq N-P \). Here, we take \( J = N-P=2 \).

Hence, multiplying both sides of Eqs. \textcolor{blue}{(\ref{Eq.A1})} and \textcolor{blue}{(\ref{Eq.A2})} by \( e^{i\omega_{-J}t} \), \( \cdots \), \( e^{i\omega_{0}t} \), \( \cdots \), \( e^{i\omega_{J}t} \) in sequence and applying the orthogonality relation in Eq. \textcolor{blue}{(\ref{Eq.A5})}, we obtain two systems of equations, each containing \( 2J+1 \) equations, expressed as:
\begin{equation}\label{Eq.A6}\tag{A6}
\begin{aligned}
& \left. \begin{aligned}
& {{R}_{-J}}=\sum\limits_{s=-J}^{+J}{\left( {{B}_{s}}\hat{u}_{s,-J}^{+}+{{C}_{s}}\hat{u}_{s,-J}^{-} \right)} \\ 
& \vdots  \\ 
& {{R}_{-1}}=\sum\limits_{s=-J}^{+J}{\left( {{B}_{s}}\hat{u}_{s,-1}^{+}+{{C}_{s}}\hat{u}_{s,-1}^{-} \right)} \\ 
& {{A}_{0}}+{{R}_{0}}=\sum\limits_{s=-J}^{+J}{\left( {{B}_{s}}\hat{u}_{s,0}^{+}+{{C}_{s}}\hat{u}_{s,0}^{-} \right)} \\ 
& {{R}_{1}}=\sum\limits_{s=-J}^{+J}{\left( {{B}_{s}}\hat{u}_{s,1}^{+}+{{C}_{s}}\hat{u}_{s,1}^{-} \right)} \\ 
& \vdots  \\ 
& {{R}_{J}}=\sum\limits_{s=-J}^{+J}{\left( {{B}_{s}}\hat{u}_{s,J}^{+}+{{C}_{s}}\hat{u}_{s,J}^{-} \right)}  
\end{aligned} \right\}(2J+1), 
\end{aligned}
\end{equation}
\begin{equation}\label{Eq.A7}\tag{A7}
\begin{aligned}
& \left. \begin{aligned}
& {{E}_{0}}\tilde{\kappa }_{-J}^{-}{{R}_{-J}}=\sum\limits_{s=-J}^{+J}{\sum\limits_{p=-P}^{+P}{{{{\hat{E}}}_{p}}\left( {{B}_{s}}\hat{\kappa }_{s,-J-p}^{+}\hat{u}_{s,-J-p}^{+}+{{C}_{s}}\hat{\kappa }_{s,-J-p}^{-}\hat{u}_{s,-J-p}^{-} \right)}} \\ 
& \vdots  \\ 
& {{E}_{0}}\tilde{\kappa }_{-1}^{-}{{R}_{-1}}=\sum\limits_{s=-J}^{+J}{\sum\limits_{p=-P}^{+P}{{{{\hat{E}}}_{p}}\left( {{B}_{s}}\hat{\kappa }_{s,-1-p}^{+}\hat{u}_{s,-1-p}^{+}+{{C}_{s}}\hat{\kappa }_{s,-1-p}^{-}\hat{u}_{s,-1-p}^{-} \right)}} \\ 
& {{E}_{0}}\left( {{\kappa }_{0}}{{A}_{0}}+\tilde{\kappa }_{0}^{-}{{R}_{0}} \right)=\sum\limits_{s=-J}^{+J}{\sum\limits_{p=-P}^{+P}{{{{\hat{E}}}_{p}}\left( {{B}_{s}}\hat{\kappa }_{s,-p}^{+}\hat{u}_{s,-p}^{+}+{{C}_{s}}\hat{\kappa }_{s,-p}^{-}\hat{u}_{s,-p}^{-} \right)}} \\ 
& {{E}_{0}}\tilde{\kappa }_{1}^{-}{{R}_{1}}=\sum\limits_{s=-J}^{+J}{\sum\limits_{p=-P}^{+P}{{{{\hat{E}}}_{p}}\left( {{B}_{s}}\hat{\kappa }_{s,1-p}^{+}\hat{u}_{s,1-p}^{+}+{{C}_{s}}\hat{\kappa }_{s,1-p}^{-}\hat{u}_{s,1-p}^{-} \right)}} \\ 
& \vdots  \\ 
& {{E}_{0}}\tilde{\kappa }_{J}^{-}{{R}_{J}}=\sum\limits_{s=-J}^{+J}{\sum\limits_{p=-P}^{+P}{{{{\hat{E}}}_{p}}\left( {{B}_{s}}\hat{\kappa }_{s,J-p}^{+}\hat{u}_{s,J-p}^{+}+{{C}_{s}}\hat{\kappa }_{s,J-p}^{-}\hat{u}_{s,J-p}^{-} \right)}}  
\end{aligned} \right\}(2J+1).
\end{aligned}
\end{equation}

Similarly, performing the same operations to Eqs. \textcolor{blue}{(\ref{Eq.A3})} and \textcolor{blue}{(\ref{Eq.A4})}, yields the following systems of equations:
\begin{equation}\label{Eq.A8}\tag{A8}
\begin{aligned}
\left. \begin{aligned}
& \sum\limits_{s=-J}^{+J}{\left( {{C}_{s}}\hat{u}_{s,-J}^{+}{{e}^{-i\hat{\kappa }_{s,-J}^{+}\Delta x}}+{{D}_{s}}\hat{u}_{s,-J}^{-}{{e}^{-i\hat{\kappa }_{s,-J}^{-}\Delta x}} \right)}={{T}_{-J}}{{e}^{-i\tilde{\kappa }_{-J}^{+}\Delta x}} \\ 
& \vdots  \\ 
& \sum\limits_{s=-J}^{+J}{\left( {{C}_{s}}\hat{u}_{s,0}^{+}{{e}^{-i\hat{\kappa }_{s,0}^{+}\Delta x}}+{{D}_{s}}\hat{u}_{s,0}^{-}{{e}^{-i\hat{\kappa }_{s,0}^{-}\Delta x}} \right)}={{T}_{0}}{{e}^{-i\tilde{\kappa }_{0}^{+}\Delta x}} \\ 
& \vdots  \\ 
& \sum\limits_{s=-J}^{+J}{\left( {{C}_{s}}\hat{u}_{s,J}^{+}{{e}^{-i\hat{\kappa }_{s,J}^{+}\Delta x}}+{{D}_{s}}\hat{u}_{s,J}^{-}{{e}^{-i\hat{\kappa }_{s,J}^{-}\Delta x}} \right)}={{T}_{J}}{{e}^{-i\tilde{\kappa }_{J}^{+}\Delta x}}  
\end{aligned} \right\}(2J+1), 
\end{aligned}
\end{equation}
\begin{equation}\label{Eq.A9}\tag{A9}
\begin{aligned}
& \left. \begin{aligned}
& \sum\limits_{s=-J}^{+J}{\sum\limits_{p=-P}^{+P}{{{{\hat{E}}}_{p}}\left( {{C}_{s}}\hat{\kappa }_{s,-J-p}^{+}\hat{u}_{s,-J-p}^{+}{{e}^{-i\hat{\kappa }_{s,-J}^{+}\Delta x}}+{{D}_{s}}\hat{\kappa }_{s,-J-p}^{-}\hat{u}_{s,-J-p}^{-}{{e}^{-i\hat{\kappa }_{s,-J}^{-}\Delta x}} \right)}} \\
& ={{E}_{0}}\tilde{\kappa }_{-J}^{+}{{T}_{-J}}{{e}^{-i\tilde{\kappa }_{-J}^{+}\Delta x}} \\ 
& \vdots  \\ 
& \sum\limits_{s=-J}^{+J}{\sum\limits_{p=-P}^{+P}{{{{\hat{E}}}_{p}}\left( {{C}_{s}}\hat{\kappa }_{s,-p}^{+}\hat{u}_{s,-p}^{+}{{e}^{-i\hat{\kappa }_{s,0}^{+}\Delta x}}+{{D}_{s}}\hat{\kappa }_{s,-p}^{-}\hat{u}_{s,-p}^{-}{{e}^{-i\hat{\kappa }_{s,0}^{-}\Delta x}} \right)}} \\
& ={{E}_{0}}\tilde{\kappa }_{0}^{+}{{T}_{0}}{{e}^{-i\tilde{\kappa }_{0}^{+}\Delta x}} \\ 
& \vdots  \\ 
& \sum\limits_{s=-J}^{+J}{\sum\limits_{p=-P}^{+P}{{{{\hat{E}}}_{p}}\left( {{C}_{s}}\hat{\kappa }_{s,J-p}^{+}\hat{u}_{s,J-p}^{+}{{e}^{-i\hat{\kappa }_{s,J}^{+}\Delta x}}+{{D}_{s}}\hat{\kappa }_{s,J-p}^{-}\hat{u}_{s,J-p}^{-}{{e}^{-i\hat{\kappa }_{s,J}^{-}\Delta x}} \right)}} \\
& ={{E}_{0}}\tilde{\kappa }_{J}^{+}{{T}_{J}}{{e}^{-i\tilde{\kappa }_{J}^{+}\Delta x}}
\end{aligned} \right\}(2J+1). 
\end{aligned}
\end{equation}

Eqs. \textcolor{blue}{(\ref{Eq.A6})}, \textcolor{blue}{(\ref{Eq.A7})} and Eqs. \textcolor{blue}{(\ref{Eq.A8})}, \textcolor{blue}{(\ref{Eq.A9})} can be reorganized respectively in matrix form as:
\begin{equation}\label{Eq.A10}\tag{A10}
{{\mathbf{M}}_{\mathbf{1}}}{{A}_{0}}+{{\mathbf{M}}_{\mathbf{2}}}\mathbf{\tilde{R}}={{\mathbf{M}}_{\mathbf{3}}}\left[ \begin{matrix}
   \mathbf{B}  \\
   \mathbf{C}  \\
\end{matrix} \right],
\end{equation}
\begin{equation}\label{Eq.A11}\tag{A11}
{{\mathbf{M}}_{\mathbf{4}}}\left[ \begin{matrix}
   \mathbf{B}  \\
   \mathbf{C}  \\
\end{matrix} \right]={{\mathbf{M}}_{\mathbf{5}}}\mathbf{\tilde{T}}.
\end{equation}
In Eqs. \textcolor{blue}{(\ref{Eq.A10})} and \textcolor{blue}{(\ref{Eq.A11})}, \( \mathbf{\tilde{R}}={{\left[ \begin{matrix}
{\tilde{R}_{-J}} & \cdots & {\tilde{R}_{0}} & \cdots  & {\tilde{R}_{J}}
\end{matrix} \right]}^{T}} \) and \( \mathbf{\tilde{T}}={{\left[ \begin{matrix}
{\tilde{T}_{-J}} & \cdots & {\tilde{T}_{0}} & \cdots  & {\tilde{T}_{J}}
\end{matrix} \right]}^{T}} \) are the amplitude vectors of reflected and transmitted waves, respectively. \( \mathbf{{B}}={{\left[ \begin{matrix}
{{B}_{-J}} & \cdots & {B_{0}} & \cdots  & {B_{J}}
\end{matrix} \right]}^{T}} \) and \( \mathbf{{C}}={{\left[ \begin{matrix}
{C_{-J}} & \cdots & {C_{0}} & \cdots  & {C_{J}}
\end{matrix} \right]}^{T}} \) are the amplitude vectors of forward and backward propagating Floquet-Bloch waves, respectively.
From Eqs. \textcolor{blue}{(\ref{Eq.A10})} and \textcolor{blue}{(\ref{Eq.A11})}, the scattering relation can thus be obtained, as shown in Eq. \textcolor{blue}{(\ref{Eq.16})}. Here, \( {\mathbf{M}}_{\mathbf{1}} \), \( {\mathbf{M}}_{\mathbf{2}} \), \( {\mathbf{M}}_{\mathbf{3}} \), \( {\mathbf{M}}_{\mathbf{4}} \) and \( {\mathbf{M}}_{\mathbf{5}} \) are transfer matrices and can be expressed as follows.

\( \mathbf{M_1} \) is a matrix of size \( (4J+2)\times1 \) and given by:
\begin{equation}\label{Eq.A12}\tag{A12}
\mathbf{M_{1}}=\begin{bmatrix}0&\cdots&1&\cdots&0&0&\cdots&E_0\kappa_{0}&\cdots&0\end{bmatrix}^{T}.
\end{equation}

\( \mathbf{M_2} \) is a matrix of size \( (4J+2)\times(2J+1) \) and given by:
\begin{equation}\label{Eq.A13}\tag{A13}
\mathbf{M_{2}}=
\begin{bmatrix}
diag(1,\;\cdots,\;1,\;\cdots,\;1)\\
diag(E_0\tilde{\kappa}_{-J}^{-},\; \cdots,\; E_0\tilde{\kappa}_{0}^{-},\; \cdots,\; E_0\tilde{\kappa}_{J}^{-})
\end{bmatrix}.
\end{equation}

\( \mathbf{M_3} \) is a matrix of size \( (4J+2)\times(4J+2) \) and given by:
\begin{small}
\begin{equation}\label{Eq.A14}\tag{A14}
\mathbf{M_3}=
\begin{bmatrix}
\hat{u}_{-J,-J}^{+} & \cdots  & \hat{u}_{J,-J}^{+} & \hat{u}_{-J,-J}^{-} & \cdots  & \hat{u}_{J,-J}^{-}  \\
\vdots  & \vdots  & \vdots  & \vdots  & \vdots  & \vdots   \\
\hat{u}_{-J,J}^{+} & \cdots  & \hat{u}_{J,J}^{+} & \hat{u}_{-J,J}^{-} & \cdots  & \hat{u}_{J,J}^{-}  \\
\sum\limits_{p=-P}^{+P}{{{{\hat{E}}}_{p}}\hat{\kappa }_{-J,-J-p}^{+}\hat{u}_{-J,-J-p}^{+}} & \cdots  & \sum\limits_{p=-P}^{+P}{{{{\hat{E}}}_{p}}\hat{\kappa }_{J,-J-p}^{+}\hat{u}_{J,-J-p}^{+}} & \sum\limits_{p=-P}^{+P}{{{{\hat{E}}}_{p}}\hat{\kappa }_{-J,-J-p}^{-}\hat{u}_{-J,-J-p}^{-}} & \cdots  & \sum\limits_{p=-P}^{+P}{{{{\hat{E}}}_{p}}\hat{\kappa }_{J,-J-p}^{-}\hat{u}_{J,-J-p}^{-}}  \\
\vdots  & \vdots  & \vdots  & \vdots  & \vdots  & \vdots   \\
\sum\limits_{p=-P}^{+P}{{{{\hat{E}}}_{p}}\hat{\kappa }_{-J,J-p}^{+}\hat{u}_{-J,J-p}^{+}} & \cdots  & \sum\limits_{p=-P}^{+P}{{{{\hat{E}}}_{p}}\hat{\kappa }_{J,J-p}^{+}\hat{u}_{J,J-p}^{+}} & \sum\limits_{p=-P}^{+P}{{{{\hat{E}}}_{p}}\hat{\kappa }_{-J,J-p}^{-}\hat{u}_{-J,J-p}^{-}} & \cdots  & \sum\limits_{p=-P}^{+P}{{{{\hat{E}}}_{p}}\hat{\kappa }_{J,J-p}^{-}\hat{u}_{J,J-p}^{-}}  \\
\end{bmatrix}.
\end{equation}
\end{small}

\( \mathbf{M_4} \) is a matrix of size \( (4J+2)\times(4J+2) \) and given by:
\begin{scriptsize}
\begin{equation}\label{Eq.A15}\tag{A15}
\mathbf{M_{4}}=
\begin{bmatrix}
\hat{u}_{-J,-J}^{+}{{e}^{-i\hat{\kappa }_{-J,-J}^{+}\Delta x}} & \cdots  & \hat{u}_{J,-J}^{+}{{e}^{-i\hat{\kappa }_{J,-J}^{+}\Delta x}} & \hat{u}_{-J,-J}^{-}{{e}^{-i\hat{\kappa }_{-J,-J}^{-}\Delta x}} & \cdots  & \hat{u}_{J,-J}^{-}{{e}^{-i\hat{\kappa }_{J,-J}^{-}\Delta x}}  \\
\vdots  & \vdots  & \vdots  & \vdots  & \vdots  & \vdots   \\
\hat{u}_{-J,J}^{+}{{e}^{-i\hat{\kappa }_{-J,J}^{+}\Delta x}} & \cdots  & \hat{u}_{J,J}^{+}{{e}^{-i\hat{\kappa }_{J,J}^{+}\Delta x}} & \hat{u}_{-J,J}^{-}{{e}^{-i\hat{\kappa }_{-J,J}^{-}\Delta x}} & \cdots  & \hat{u}_{J,J}^{-}{{e}^{-i\hat{\kappa }_{J,J}^{-}\Delta x}}  \\
\sum\limits_{p=-P}^{P}{{{{\hat{E}}}_{p}}\hat{\kappa }_{-J,-J-p}^{+}\hat{u}_{-J,-J-p}^{+}{{e}^{-i\hat{\kappa }_{-J,-J}^{-}\Delta x}}} & \cdots  & \sum\limits_{p=-P}^{P}{{{{\hat{E}}}_{p}}\hat{\kappa }_{J,-J-p}^{+}\hat{u}_{J,-J-p}^{+}{{e}^{-i\hat{\kappa }_{J,-J}^{+}\Delta x}}} & \sum\limits_{p=-P}^{P}{{{{\hat{E}}}_{p}}\hat{\kappa }_{-J,-J-p}^{-}\hat{u}_{-J,-J-p}^{-}{{e}^{-i\hat{\kappa }_{-J,-J}^{-}\Delta x}}} & \cdots  & \sum\limits_{p=-P}^{P}{{{{\hat{E}}}_{p}}\hat{\kappa }_{J,-J-p}^{-}\hat{u}_{J,-J-p}^{-}{{e}^{-i\hat{\kappa }_{J,-J}^{-}\Delta x}}}  \\
\vdots  & \vdots  & \vdots  & \vdots  & \vdots  & \vdots   \\
\sum\limits_{p=-P}^{P}{{{{\hat{E}}}_{p}}\hat{\kappa }_{-J,J-p}^{+}\hat{u}_{-J,J-p}^{+}{{e}^{-i\hat{\kappa }_{-J,J}^{+}\Delta x}}} & \cdots  & \sum\limits_{p=-P}^{P}{{{{\hat{E}}}_{p}}\hat{\kappa }_{J,J-p}^{+}\hat{u}_{J,J-p}^{+}{{e}^{-i\hat{\kappa }_{J,J}^{+}\Delta x}}} & \sum\limits_{p=-P}^{P}{{{{\hat{E}}}_{p}}\hat{\kappa }_{-J,J-p}^{-}\hat{u}_{-J,J-p}^{-}{{e}^{-i\hat{\kappa }_{-J,J}^{-}\Delta x}}} & \cdots  & \sum\limits_{p=-P}^{P}{{{{\hat{E}}}_{p}}\hat{\kappa }_{J,J-p}^{-}\hat{u}_{J,J-p}^{-}{{e}^{-i\hat{\kappa }_{J,J}^{-}\Delta x}}}  \\
\end{bmatrix}.
\end{equation}
\end{scriptsize}

\( \mathbf{M_5} \) is a matrix of size \( (4J+2)\times(2J+1) \) and given by:
\begin{equation}\label{Eq.A16}\tag{A16}
\mathbf{M_{5}}=
\begin{bmatrix}
diag(e^{-i\tilde{\kappa}_{-J}^{+}\Delta x},\; \cdots,\; e^{-i\tilde{\kappa}_{0}^{+}\Delta x},\; \cdots,\; e^{-i\tilde{\kappa}_{J}^{+}\Delta x})\\
diag(E_0\tilde{\kappa}_{-J}^{+}e^{-i\tilde{\kappa}_{-J}^{+}\Delta x},\; \cdots,\; E_0\tilde{\kappa}_{0}^{+}e^{-i\tilde{\kappa}_{0}^{+}\Delta x},\; \cdots,\; E_0\tilde{\kappa}_{J}^{+}e^{-i\tilde{\kappa}_{J}^{+}\Delta x})
\end{bmatrix}.
\end{equation}

By substituting Eqs. \textcolor{blue}{(\ref{Eq.A12})}-\textcolor{blue}{(\ref{Eq.A16})} into Eq. \textcolor{blue}{(\ref{Eq.16})}, the transmission and reflection coefficients from the \( -J^\mathrm{th} \)-order to the \( J^\mathrm{th} \)-order can be obtained.

\section{Tuning the operating frequency of parametric amplification via the dimensionless modulation velocity \( V \)}
\label{App.B}
\renewcommand{\thefigure}{B.\arabic{figure}}
\setcounter{figure}{0}
In this appendix, we investigate the effect of the dimensionless modulation velocity \( V \) on scattering coefficients for supersonic modulation. For simplicity, we consider positive incidence as a representative case to illustrate the influence of dimensionless modulation velocity \( V \) on the parametric amplification effect. From the previous analysis in the main text, only the \( 0^\mathrm{th} \)-order transmission and \( -1^\mathrm{st} \)-order reflection modes are amplified, while other scattering modes always remain at a lower amplitude. Hence, we focus on the evolution of the \( 0^\mathrm{th} \)-order transmission and \( -1^\mathrm{st} \)-order reflection profiles with varying dimensionless modulation velocity \( V \), as shown in \textcolor{blue}{Figs. \ref{Fig.B.1}}(a) and (b). It can be observed that both the \( 0^\mathrm{th} \)-order transmission and \( -1^\mathrm{st} \)-order reflection peaks shift toward higher frequencies as \( V \) increases, consistently aligning with the center frequency of the forward wavenumber bandgap, indicated by the black dashed lines (i.e., \( \Omega_0=\frac{1+V}{2} \)). Therefore, we can easily customize parametric amplification and frequency conversion effects for different operation frequencies by adjusting the dimensionless modulation velocity \( V \). In particular, the parametric amplification effect is strongest at \( V=2.45 \) within the range of interest, with maximum amplified peak values of \( T_{0,peak} = 37.2 \) and \( R_{-1,peak} = 58 \), indicating that the proposed supersonic spatiotemporal modulation design exhibits excellent parametric amplification capability.

\begin{figure}[h]   \centering
\includegraphics[width=0.75\linewidth]{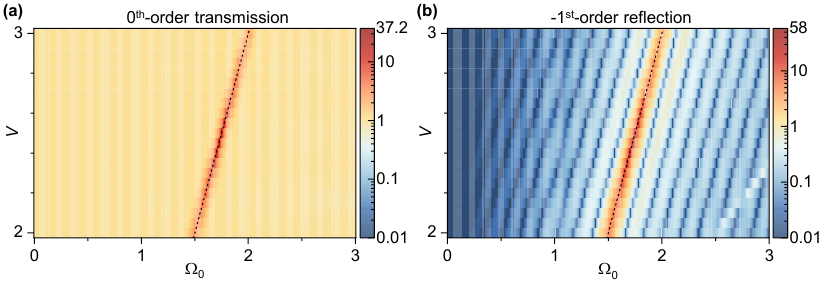}
\caption{Parametric analysis of dimensionless modulation velocity \( V \) for positive incidence. (a) \( 0^\mathrm{th} \)-order transmission and (b) \( -1^\mathrm{st} \)-order reflection coefficient profiles varying as dimensionless modulation velocity \( V \). The black dashed line tracks the center frequency shift of the forward wavenumber bandgap with increasing \( V \).}
\label{Fig.B.1}
\end{figure}

\section{Parametric instability under large modulation length and modulation amplitude}
\label{App.C}
\renewcommand{\thefigure}{C.\arabic{figure}}
\setcounter{figure}{0}

\begin{figure}[h]   \centering
\includegraphics[width=0.9\linewidth]{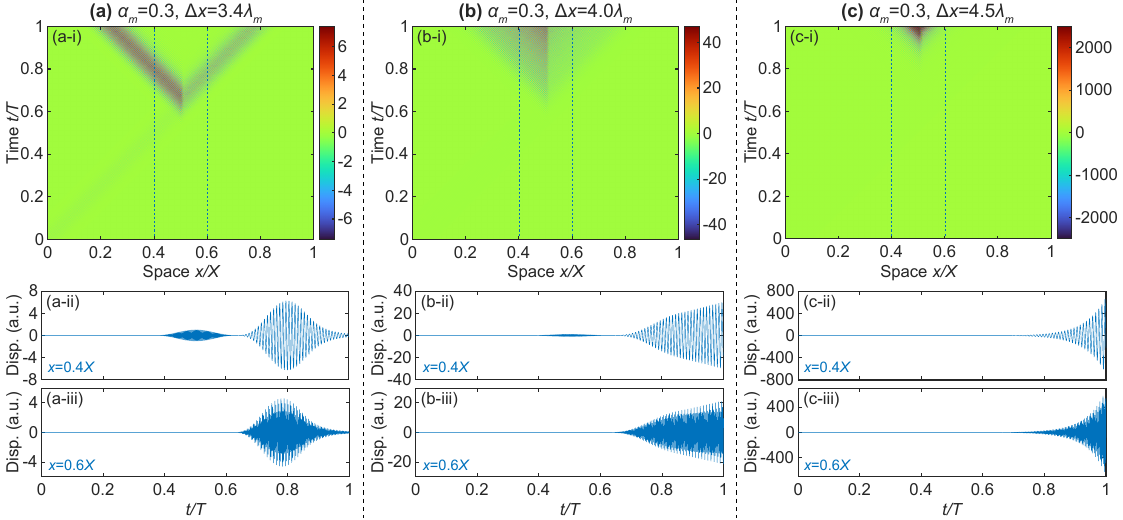}
\caption{Numerical demonstration of the transition from parametric stability to parametric instability induced by excessive modulation length. (a) Parametric stability: \( \alpha_m = 0.3 \), \( \Delta x = 3.4\lambda_m \). (b) Parametric instability: \( \alpha_m = 0.3 \), \( \Delta x = 4.0\lambda_m \). (c) Parametric instability: \( \alpha_m = 0.3 \), \( \Delta x = 4.5\lambda_m \). (a/b/c-i) Time-history displacement wavefield at \( \Omega_0 = 1.5 \). Corresponding displacement responses at (a/b/c-ii) \( x = 0.4X \) (reflection region) and (a/b/c-iii) \( x = 0.6X \) (transmission region).}
\label{Fig.C.1}
\end{figure}

\begin{figure}[h]   \centering
\includegraphics[width=0.9\linewidth]{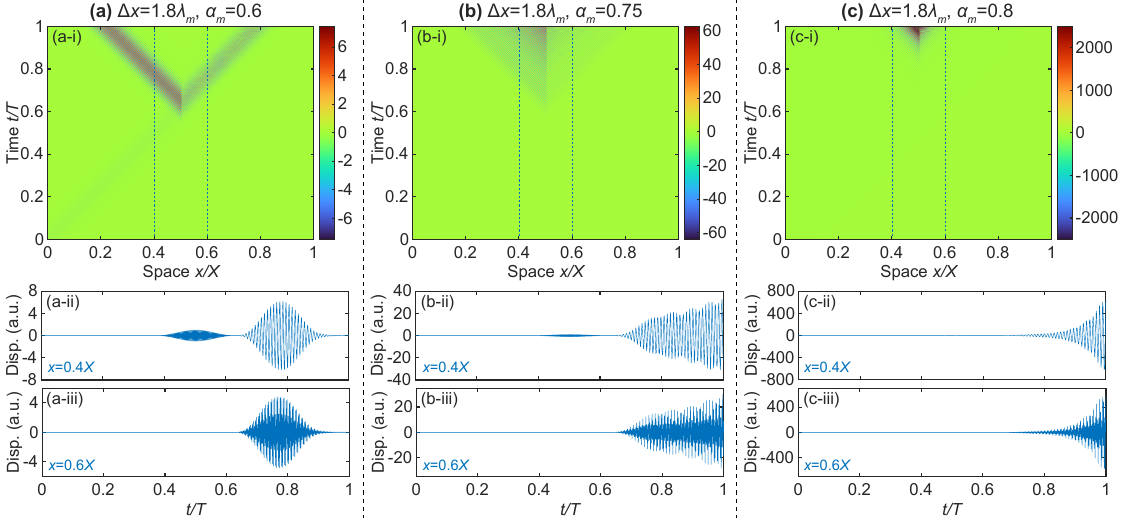}
\caption{Numerical demonstration of the transition from parametric stability to parametric instability induced by excessive modulation amplitude. (a) Parametric stability: \( \Delta x = 1.8\lambda_m \), \( \alpha_m = 0.6 \). (b) Parametric instability: \( \Delta x = 1.8\lambda_m \), \( \alpha_m = 0.75 \). (c) Parametric instability: \( \Delta x = 1.8\lambda_m \), \( \alpha_m = 0.8 \). (a/b/c-i) Time-history displacement wavefield at \( \Omega_0 = 1.5 \). Corresponding displacement responses at (a/b/c-ii) \( x = 0.4X \) (reflection region) and (a/b/c-iii) \( x = 0.6X \) (transmission region).}
\label{Fig.C.2}
\end{figure}

In this appendix, we numerically demonstrate parametric instability arising from large modulation length and modulation amplitude for a representative frequency \( \Omega_0 = 1.5 \). We begin by examining the effect of excessively increased modulation length. For reference, the system’s stable displacement response is shown in \textcolor{blue}{Fig. \ref{Fig.C.1}}(a). Displacement responses within the reflection (\( x = 0.4X \)) and transmission (\( x = 0.6X \)) regions are extracted and presented in \textcolor{blue}{Figs. \ref{Fig.C.1}}(a-ii)--(a-iii). These results show stable, amplified reflected and transmitted wave packets, characterized by bounded displacement responses that decay back to zero over time. For excessive modulation length, the system exhibits an unbounded displacement growth over time, indicating parametric instability, as shown in \textcolor{blue}{Figs. \ref{Fig.C.1}}(b)--(c). Moreover, the rate of this unbounded growth increases with the modulation length. A similar transition from stable to unstable response is observed as the modulation amplitude increases excessively, as shown in \textcolor{blue}{Figs. \ref{Fig.C.2}}(a)--(c). In the unstable regime, larger modulation amplitudes lead to faster unbounded displacement growth over time. It is important to note that the theoretical framework for supersonic modulation proposed in this work applies only to parameter-stable systems, thereby ensuring the physical validity of the derived scattering coefficients. Future work will address the analysis of parametric instability in greater detail.

\section{Numerical demonstration for nonreciprocal performance}
\label{App.D}
\renewcommand{\thefigure}{D.\arabic{figure}}
\setcounter{figure}{0}
As an illustration, we focus only on the forward \( \mu \)-bandgap located at \( \Omega=1.5 \) to demonstrate nonreciprocal wave propagation. \textcolor{blue}{Fig. \ref{Fig.D.1}}(a) shows the normalized FFT spectra of incident, transmitted, and reflected waves for positive incidence. We further obtain the wavefield evolution in the wavenumber-frequency domain by extracting wavenumber and frequency properties in different time windows, i.e., \( [115\lambda_m/c_0,125\lambda_m/c_0] \), \( [120\lambda_m/c_0,130\lambda_m/c_0] \), \( [125\lambda_m/c_0,135\lambda_m/c_0] \), and \( [130\lambda_m/c_0,140\lambda_m/c_0] \), as presented in \textcolor{blue}{Figs. \ref{Fig.D.1}}(b-i)-(b-iv). Recalling that modulation is activated and deactivated at \( x_0=125\lambda_m \) and \( x_1=128\lambda_m \), \textcolor{blue}{Figs. \ref{Fig.D.1}}(b-i) and (b-iv) correspond to the wavenumber-frequency distribution of incident and scattered wavefields, while \textcolor{blue}{Figs. \ref{Fig.D.1}}(b-ii) and (b-iii) display the energy transfer and accumulation as modulation takes effect. To aid in understanding, the dispersion diagram of the modulated media is also shown as background gray lines. It is evident that for a positive incident wave centered in the forward \( \mu \)-bandgap, supersonic modulation excites a frequency-converted reflection mode within the backward \( \mu \)-bandgap and amplifies the transmitted and reflected wave energy near the \( \mu \)-bandgaps. However, for the negative incidence case shown in \textcolor{blue}{Fig. \ref{Fig.D.1}}(c), the transmission spectrum remains identical to the incident spectrum, with nearly no observable reflection. The wavefield evolution in the wavenumber-frequency domain, as depicted in \textcolor{blue}{Figs. \ref{Fig.D.1}}(d-i)-(d-iv), further confirms the absence of parametric amplification and frequency conversion. This demonstrates that nonreciprocal parametric amplification and frequency conversion are realized within the wavenumber bandgap induced by a supersonic modulated medium.

\begin{figure}[h]   \centering
\includegraphics[width=0.9\linewidth]{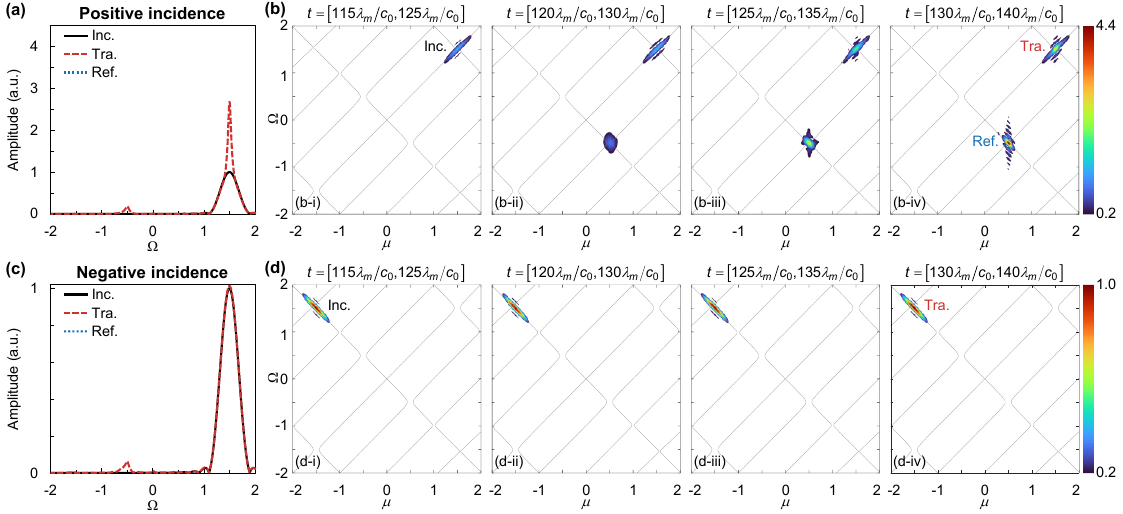}
\caption{Numerical demonstration of nonreciprocal parametric amplification and frequency conversion for the forward \( \mu \)-bandgap at \( \Omega=1.5 \). (a) Positive incidence: normalized FFT spectra of the incident (black solid line), transmitted (red dashed line), and reflected (blue dotted line) waves. (b) Positive incidence: wavefield evolution in the wavenumber-frequency domain for different time windows (b-i) \( [115\lambda_m/c_0,125\lambda_m/c_0] \), (b-ii) \( [120\lambda_m/c_0,130\lambda_m/c_0] \), (b-iii) \( [125\lambda_m/c_0,135\lambda_m/c_0] \), and (b-iv) \( [130\lambda_m/c_0,140\lambda_m/c_0] \). (c) Negative incidence: normalized FFT spectra of the incident (black solid line), transmitted (red dashed line), and reflected (blue dotted line) waves. (d) Negative incidence: wavefield evolution in the wavenumber-frequency domain for different time windows (d-i) \( [115\lambda_m/c_0,125\lambda_m/c_0] \), (d-ii) \( [120\lambda_m/c_0,130\lambda_m/c_0] \), (d-iii) \( [125\lambda_m/c_0,135\lambda_m/c_0] \), and (d-iv) \( [130\lambda_m/c_0,140\lambda_m/c_0] \).}
\label{Fig.D.1}
\end{figure}

\section{Numerical demonstration of the influence of modulation length and modulation amplitude}
\label{App.E}
\renewcommand{\thefigure}{E.\arabic{figure}}
\setcounter{figure}{0}
The effect of modulation parameters on amplification performance is further verified through numerical simulations, with results shown only for positive incidence to avoid redundancy. Numerical scattering coefficients of the \( 0^\mathrm{th} \)-order transmission and \( -1^\mathrm{st} \)-order reflection for various modulation lengths are shown in \textcolor{blue}{Figs. \ref{Fig.E1}}(a) and (b), demonstrating good agreement with the analytical results. The corresponding time-history amplitude wavefields at \( \Omega_0 = 1.5 \) for different modulation lengths are extracted and presented in \textcolor{blue}{Fig. \ref{Fig.E1}}(c). As the modulation length increases, the intensity of both transmitted and reflected wavefields gradually improves, confirming the role of modulation length in enhancing amplification.

\begin{figure}[h]   \centering
\includegraphics[width=0.9\linewidth]{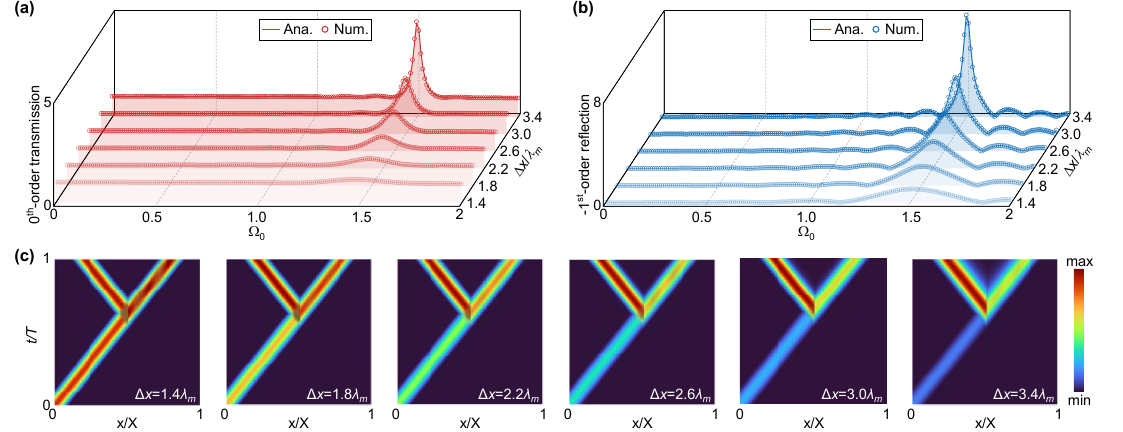}
\caption{Numerical demonstration of the influence of modulation length on amplification performance, with fixed parameters: \( \kappa_m=10\;\mathrm{rad/m} \), \( \omega_m=20\;\mathrm{rad/s} \), \( \alpha_m=0.3 \). Scattering coefficients of the (a) \( 0^\mathrm{th} \)-order transmission and (b) \( -1^\mathrm{st} \)-order reflection for positive incidence. (c) Time-history amplitude wavefields at \( \Omega_0 = 1.5 \) for various modulation lengths.}
\label{Fig.E1}
\end{figure}

\begin{figure}[h]   \centering
\includegraphics[width=0.9\linewidth]{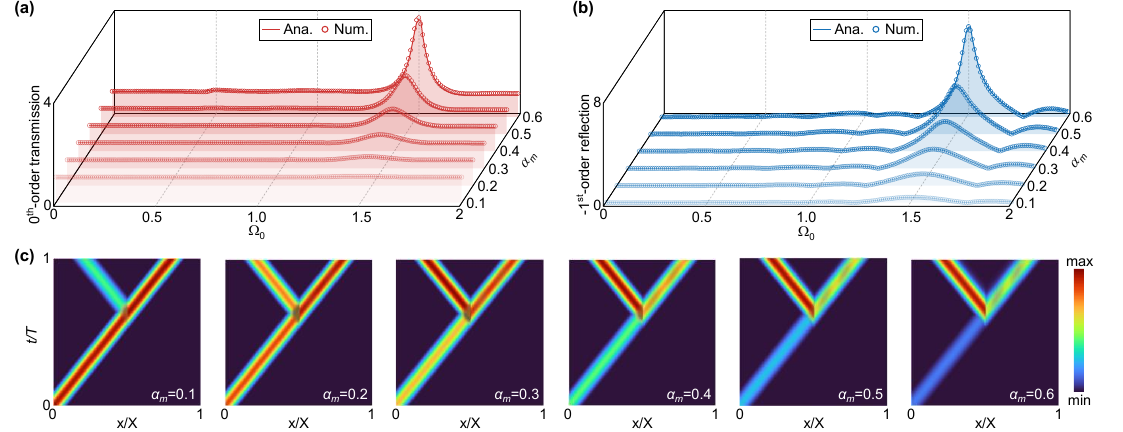}
\caption{Numerical demonstration of the influence of modulation amplitude on amplification performance, with fixed parameters: \( \kappa_m=10\;\mathrm{rad/m} \), \( \omega_m=20\;\mathrm{rad/s} \), \( \Delta x=1.8\lambda_m \). Scattering coefficients of the (a) \( 0^\mathrm{th} \)-order transmission and (b) \( -1^\mathrm{st} \)-order reflection for positive incidence. (c) Time-history amplitude wavefields at \( \Omega_0 = 1.5 \) for various modulation amplitudes.}
\label{Fig.E2}
\end{figure}

In addition, the influence of modulation amplitude is also numerically verified. \textcolor{blue}{Figs. \ref{Fig.E2}}(a) and (b) show the dominant transmission and reflection coefficients for various modulation amplitudes, which also exhibit good agreement with the analytical predictions. As expected, an enhancement of the scattered wavefields is clearly observed at \( \Omega_0 = 1.5 \) with increasing modulation amplitude, as illustrated in \textcolor{blue}{Fig. \ref{Fig.E2}}(c).

\printcredits

\bibliographystyle{cas-model2-names}

\bibliography{cas-refs}

\begin{thebibliography}{40}
\expandafter\ifx\csname natexlab\endcsname\relax\def\natexlab#1{#1}\fi
\providecommand{\url}[1]{\texttt{#1}}
\providecommand{\href}[2]{#2}
\providecommand{\path}[1]{#1}
\providecommand{\DOIprefix}{doi:}
\providecommand{\ArXivprefix}{arXiv:}
\providecommand{\URLprefix}{URL: }
\providecommand{\Pubmedprefix}{pmid:}
\providecommand{\doi}[1]{\href{http://dx.doi.org/#1}{\path{#1}}}
\providecommand{\Pubmed}[1]{\href{pmid:#1}{\path{#1}}}
\providecommand{\bibinfo}[2]{#2}
\ifx\xfnm\relax \def\xfnm[#1]{\unskip,\space#1}\fi
%Type = Article
\bibitem[{Sounas and Al{\`u}(2017)}]{sounas2017non}
\bibinfo{author}{Sounas, D.L.}, \bibinfo{author}{Al{\`u}, A.}, \bibinfo{year}{2017}.
\newblock \bibinfo{title}{Non-reciprocal photonics based on time modulation}.
\newblock \bibinfo{journal}{Nature Photonics} \bibinfo{volume}{11}, \bibinfo{pages}{774--783}.
%Type = Article
\bibitem[{Nassar et~al.(2020)Nassar, Yousefzadeh, Fleury, Ruzzene, Al{\`u}, Daraio, Norris, Huang and Haberman}]{nassar2020nonreciprocity}
\bibinfo{author}{Nassar, H.}, \bibinfo{author}{Yousefzadeh, B.}, \bibinfo{author}{Fleury, R.}, \bibinfo{author}{Ruzzene, M.}, \bibinfo{author}{Al{\`u}, A.}, \bibinfo{author}{Daraio, C.}, \bibinfo{author}{Norris, A.N.}, \bibinfo{author}{Huang, G.}, \bibinfo{author}{Haberman, M.R.}, \bibinfo{year}{2020}.
\newblock \bibinfo{title}{Nonreciprocity in acoustic and elastic materials}.
\newblock \bibinfo{journal}{Nature Reviews Materials} \bibinfo{volume}{5}, \bibinfo{pages}{667--685}.
%Type = Article
\bibitem[{Nagulu et~al.(2020)Nagulu, Reiskarimian and Krishnaswamy}]{nagulu2020non}
\bibinfo{author}{Nagulu, A.}, \bibinfo{author}{Reiskarimian, N.}, \bibinfo{author}{Krishnaswamy, H.}, \bibinfo{year}{2020}.
\newblock \bibinfo{title}{Non-reciprocal electronics based on temporal modulation}.
\newblock \bibinfo{journal}{Nature Electronics} \bibinfo{volume}{3}, \bibinfo{pages}{241--250}.
%Type = Article
\bibitem[{Kutsaev et~al.(2021)Kutsaev, Krasnok, Romanenko, Smirnov, Taletski and Yakovlev}]{kutsaev2021up}
\bibinfo{author}{Kutsaev, S.V.}, \bibinfo{author}{Krasnok, A.}, \bibinfo{author}{Romanenko, S.N.}, \bibinfo{author}{Smirnov, A.Y.}, \bibinfo{author}{Taletski, K.}, \bibinfo{author}{Yakovlev, V.P.}, \bibinfo{year}{2021}.
\newblock \bibinfo{title}{Up-and-coming advances in optical and microwave nonreciprocity: From classical to quantum realm}.
\newblock \bibinfo{journal}{Advanced Photonics Research} \bibinfo{volume}{2}, \bibinfo{pages}{2000104}.
%Type = Article
\bibitem[{Wang et~al.(2023)Wang, Li, Wang, Sano, Sun, Shao, Takeishi, Matsubara, Okumura, Sakai et~al.}]{wang2023mechanical}
\bibinfo{author}{Wang, X.}, \bibinfo{author}{Li, Z.}, \bibinfo{author}{Wang, S.}, \bibinfo{author}{Sano, K.}, \bibinfo{author}{Sun, Z.}, \bibinfo{author}{Shao, Z.}, \bibinfo{author}{Takeishi, A.}, \bibinfo{author}{Matsubara, S.}, \bibinfo{author}{Okumura, D.}, \bibinfo{author}{Sakai, N.}, et~al., \bibinfo{year}{2023}.
\newblock \bibinfo{title}{Mechanical nonreciprocity in a uniform composite material}.
\newblock \bibinfo{journal}{Science} \bibinfo{volume}{380}, \bibinfo{pages}{192--198}.
%Type = Article
\bibitem[{Yang et~al.(2024)Yang, Liu, Zhao, Fan and Qiu}]{yang2024nonreciprocal}
\bibinfo{author}{Yang, S.}, \bibinfo{author}{Liu, M.}, \bibinfo{author}{Zhao, C.}, \bibinfo{author}{Fan, S.}, \bibinfo{author}{Qiu, C.W.}, \bibinfo{year}{2024}.
\newblock \bibinfo{title}{Nonreciprocal thermal photonics}.
\newblock \bibinfo{journal}{Nature photonics} \bibinfo{volume}{18}, \bibinfo{pages}{412--424}.
%Type = Article
\bibitem[{Liang et~al.(2010)Liang, Guo, Tu, Zhang and Cheng}]{liang2010acoustic}
\bibinfo{author}{Liang, B.}, \bibinfo{author}{Guo, X.}, \bibinfo{author}{Tu, J.}, \bibinfo{author}{Zhang, D.}, \bibinfo{author}{Cheng, J.}, \bibinfo{year}{2010}.
\newblock \bibinfo{title}{An acoustic rectifier}.
\newblock \bibinfo{journal}{Nature materials} \bibinfo{volume}{9}, \bibinfo{pages}{989--992}.
%Type = Article
\bibitem[{Fleury et~al.(2014)Fleury, Sounas, Sieck, Haberman and Al{\`u}}]{fleury2014sound}
\bibinfo{author}{Fleury, R.}, \bibinfo{author}{Sounas, D.L.}, \bibinfo{author}{Sieck, C.F.}, \bibinfo{author}{Haberman, M.R.}, \bibinfo{author}{Al{\`u}, A.}, \bibinfo{year}{2014}.
\newblock \bibinfo{title}{Sound isolation and giant linear nonreciprocity in a compact acoustic circulator}.
\newblock \bibinfo{journal}{Science} \bibinfo{volume}{343}, \bibinfo{pages}{516--519}.
%Type = Article
\bibitem[{Zhang et~al.(2024)Zhang, Ge, Guan, Chen, Han, Chen, Pan, Jia, Yuan, Sun et~al.}]{zhang2024nonreciprocal}
\bibinfo{author}{Zhang, L.}, \bibinfo{author}{Ge, Y.}, \bibinfo{author}{Guan, Y.j.}, \bibinfo{author}{Chen, F.}, \bibinfo{author}{Han, N.}, \bibinfo{author}{Chen, Q.}, \bibinfo{author}{Pan, Y.}, \bibinfo{author}{Jia, D.}, \bibinfo{author}{Yuan, S.q.}, \bibinfo{author}{Sun, H.x.}, et~al., \bibinfo{year}{2024}.
\newblock \bibinfo{title}{Nonreciprocal acoustic devices with asymmetric peierls phases}.
\newblock \bibinfo{journal}{Physical Review Letters} \bibinfo{volume}{133}, \bibinfo{pages}{136601}.
%Type = Article
\bibitem[{Wang et~al.(2013)Wang, Zhou, Guo, Zhang, Evers and Zhu}]{wang2013optical}
\bibinfo{author}{Wang, D.W.}, \bibinfo{author}{Zhou, H.T.}, \bibinfo{author}{Guo, M.J.}, \bibinfo{author}{Zhang, J.X.}, \bibinfo{author}{Evers, J.}, \bibinfo{author}{Zhu, S.Y.}, \bibinfo{year}{2013}.
\newblock \bibinfo{title}{Optical diode made from a moving photonic crystal}.
\newblock \bibinfo{journal}{Phys. Rev. Lett.} \bibinfo{volume}{110}, \bibinfo{pages}{093901}.
%Type = Article
\bibitem[{Li et~al.(2011)Li, Ni, Feng, Lu, He and Chen}]{li2011tunable}
\bibinfo{author}{Li, X.F.}, \bibinfo{author}{Ni, X.}, \bibinfo{author}{Feng, L.}, \bibinfo{author}{Lu, M.H.}, \bibinfo{author}{He, C.}, \bibinfo{author}{Chen, Y.F.}, \bibinfo{year}{2011}.
\newblock \bibinfo{title}{Tunable unidirectional sound propagation through a sonic-crystal-based acoustic diode}.
\newblock \bibinfo{journal}{Physical review letters} \bibinfo{volume}{106}, \bibinfo{pages}{084301}.
%Type = Article
\bibitem[{Jal{\v{s}}i{\'c} et~al.(2023)Jal{\v{s}}i{\'c}, Alujevi{\'c}, Garma, {\'C}atipovi{\'c}, Joki{\'c} and Wolf}]{jalvsic2023active}
\bibinfo{author}{Jal{\v{s}}i{\'c}, M.}, \bibinfo{author}{Alujevi{\'c}, N.}, \bibinfo{author}{Garma, T.}, \bibinfo{author}{{\'C}atipovi{\'c}, I.}, \bibinfo{author}{Joki{\'c}, M.}, \bibinfo{author}{Wolf, H.}, \bibinfo{year}{2023}.
\newblock \bibinfo{title}{An active metamaterial cell concept for nonreciprocal vibroacoustic transmission}.
\newblock \bibinfo{journal}{Mechanical systems and signal processing} \bibinfo{volume}{186}, \bibinfo{pages}{109829}.
%Type = Article
\bibitem[{Koutserimpas and Fleury(2018)}]{koutserimpas2018nonreciprocal}
\bibinfo{author}{Koutserimpas, T.T.}, \bibinfo{author}{Fleury, R.}, \bibinfo{year}{2018}.
\newblock \bibinfo{title}{Nonreciprocal gain in non-hermitian time-floquet systems}.
\newblock \bibinfo{journal}{Physical Review Letters} \bibinfo{volume}{120}, \bibinfo{pages}{087401}.
%Type = Article
\bibitem[{Shen et~al.(2023)Shen, Zhang, Chen, Xiao, Zou, Guo and Dong}]{shen2023nonreciprocal}
\bibinfo{author}{Shen, Z.}, \bibinfo{author}{Zhang, Y.L.}, \bibinfo{author}{Chen, Y.}, \bibinfo{author}{Xiao, Y.F.}, \bibinfo{author}{Zou, C.L.}, \bibinfo{author}{Guo, G.C.}, \bibinfo{author}{Dong, C.H.}, \bibinfo{year}{2023}.
\newblock \bibinfo{title}{Nonreciprocal frequency conversion and mode routing in a microresonator}.
\newblock \bibinfo{journal}{Physical Review Letters} \bibinfo{volume}{130}, \bibinfo{pages}{013601}.
%Type = Article
\bibitem[{Guo et~al.(2023)Guo, Lissek and Fleury}]{guo2023observation}
\bibinfo{author}{Guo, X.}, \bibinfo{author}{Lissek, H.}, \bibinfo{author}{Fleury, R.}, \bibinfo{year}{2023}.
\newblock \bibinfo{title}{Observation of non-reciprocal harmonic conversion in real sounds}.
\newblock \bibinfo{journal}{Communications Physics} \bibinfo{volume}{6}, \bibinfo{pages}{93}.
%Type = Article
\bibitem[{Fleury et~al.(2016)Fleury, Khanikaev and Alu}]{fleury2016floquet}
\bibinfo{author}{Fleury, R.}, \bibinfo{author}{Khanikaev, A.B.}, \bibinfo{author}{Alu, A.}, \bibinfo{year}{2016}.
\newblock \bibinfo{title}{Floquet topological insulators for sound}.
\newblock \bibinfo{journal}{Nature communications} \bibinfo{volume}{7}, \bibinfo{pages}{11744}.
%Type = Article
\bibitem[{Tian et~al.(2022)Tian, Zhang, Zhang, Wu, Lin, Zhou, Duan, Jiang and Du}]{tian2022experimental}
\bibinfo{author}{Tian, T.}, \bibinfo{author}{Zhang, Y.}, \bibinfo{author}{Zhang, L.}, \bibinfo{author}{Wu, L.}, \bibinfo{author}{Lin, S.}, \bibinfo{author}{Zhou, J.}, \bibinfo{author}{Duan, C.K.}, \bibinfo{author}{Jiang, J.H.}, \bibinfo{author}{Du, J.}, \bibinfo{year}{2022}.
\newblock \bibinfo{title}{Experimental realization of nonreciprocal adiabatic transfer of phonons in a dynamically modulated nanomechanical topological insulator}.
\newblock \bibinfo{journal}{Physical Review Letters} \bibinfo{volume}{129}, \bibinfo{pages}{215901}.
%Type = Article
\bibitem[{Virtanen and Heikkil{\"a}(2024)}]{virtanen2024nonreciprocal}
\bibinfo{author}{Virtanen, P.}, \bibinfo{author}{Heikkil{\"a}, T.T.}, \bibinfo{year}{2024}.
\newblock \bibinfo{title}{Nonreciprocal josephson linear response}.
\newblock \bibinfo{journal}{Physical Review Letters} \bibinfo{volume}{132}, \bibinfo{pages}{046002}.
%Type = Article
\bibitem[{Lepri and Casati(2011)}]{lepri2011asymmetric}
\bibinfo{author}{Lepri, S.}, \bibinfo{author}{Casati, G.}, \bibinfo{year}{2011}.
\newblock \bibinfo{title}{Asymmetric wave propagation in nonlinear systems}.
\newblock \bibinfo{journal}{Physical review letters} \bibinfo{volume}{106}, \bibinfo{pages}{164101}.
%Type = Article
\bibitem[{Cotrufo et~al.(2024)Cotrufo, Cordaro, Sounas, Polman and Al{\`u}}]{cotrufo2024passive}
\bibinfo{author}{Cotrufo, M.}, \bibinfo{author}{Cordaro, A.}, \bibinfo{author}{Sounas, D.L.}, \bibinfo{author}{Polman, A.}, \bibinfo{author}{Al{\`u}, A.}, \bibinfo{year}{2024}.
\newblock \bibinfo{title}{Passive bias-free non-reciprocal metasurfaces based on thermally nonlinear quasi-bound states in the continuum}.
\newblock \bibinfo{journal}{Nature Photonics} \bibinfo{volume}{18}, \bibinfo{pages}{81--90}.
%Type = Article
\bibitem[{Ma et~al.(2019)Ma, Xiao and Chan}]{ma2019topological}
\bibinfo{author}{Ma, G.}, \bibinfo{author}{Xiao, M.}, \bibinfo{author}{Chan, C.T.}, \bibinfo{year}{2019}.
\newblock \bibinfo{title}{Topological phases in acoustic and mechanical systems}.
\newblock \bibinfo{journal}{Nature Reviews Physics} \bibinfo{volume}{1}, \bibinfo{pages}{281--294}.
%Type = Article
\bibitem[{Wang et~al.(2021)Wang, Zhang, Zhang, Wang, Guo, Zhang and Chan}]{wang2021topological_2}
\bibinfo{author}{Wang, M.}, \bibinfo{author}{Zhang, R.Y.}, \bibinfo{author}{Zhang, L.}, \bibinfo{author}{Wang, D.}, \bibinfo{author}{Guo, Q.}, \bibinfo{author}{Zhang, Z.Q.}, \bibinfo{author}{Chan, C.T.}, \bibinfo{year}{2021}.
\newblock \bibinfo{title}{Topological one-way large-area waveguide states in magnetic photonic crystals}.
\newblock \bibinfo{journal}{Physical Review Letters} \bibinfo{volume}{126}, \bibinfo{pages}{067401}.
%Type = Article
\bibitem[{Nassar et~al.(2017)Nassar, Xu, Norris and Huang}]{nassar2017modulated}
\bibinfo{author}{Nassar, H.}, \bibinfo{author}{Xu, X.}, \bibinfo{author}{Norris, A.}, \bibinfo{author}{Huang, G.}, \bibinfo{year}{2017}.
\newblock \bibinfo{title}{Modulated phononic crystals: Non-reciprocal wave propagation and willis materials}.
\newblock \bibinfo{journal}{Journal of the Mechanics and Physics of Solids} \bibinfo{volume}{101}, \bibinfo{pages}{10\--29}.
%Type = Article
\bibitem[{Galiffi et~al.(2022)Galiffi, Tirole, Yin, Li, Vezzoli, Huidobro, Silveirinha, Sapienza, Al{\`u} and Pendry}]{galiffi2022photonics}
\bibinfo{author}{Galiffi, E.}, \bibinfo{author}{Tirole, R.}, \bibinfo{author}{Yin, S.}, \bibinfo{author}{Li, H.}, \bibinfo{author}{Vezzoli, S.}, \bibinfo{author}{Huidobro, P.A.}, \bibinfo{author}{Silveirinha, M.G.}, \bibinfo{author}{Sapienza, R.}, \bibinfo{author}{Al{\`u}, A.}, \bibinfo{author}{Pendry, J.B.}, \bibinfo{year}{2022}.
\newblock \bibinfo{title}{Photonics of time-varying media}.
\newblock \bibinfo{journal}{Advanced Photonics} \bibinfo{volume}{4}, \bibinfo{pages}{014002--014002}.
%Type = Article
\bibitem[{Trainiti et~al.(2019)Trainiti, Xia, Marconi, Cazzulani, Erturk and Ruzzene}]{Trainiti2019time}
\bibinfo{author}{Trainiti, G.}, \bibinfo{author}{Xia, Y.}, \bibinfo{author}{Marconi, J.}, \bibinfo{author}{Cazzulani, G.}, \bibinfo{author}{Erturk, A.}, \bibinfo{author}{Ruzzene, M.}, \bibinfo{year}{2019}.
\newblock \bibinfo{title}{Time-periodic stiffness modulation in elastic metamaterials for selective wave filtering: Theory and experiment}.
\newblock \bibinfo{journal}{Physical Review Letters} \bibinfo{volume}{122}, \bibinfo{pages}{124301}.
%Type = Article
\bibitem[{Marconi et~al.(2020)Marconi, Riva, Di~Ronco, Cazzulani, Braghin and Ruzzene}]{Marconi2020experimental}
\bibinfo{author}{Marconi, J.}, \bibinfo{author}{Riva, E.}, \bibinfo{author}{Di~Ronco, M.}, \bibinfo{author}{Cazzulani, G.}, \bibinfo{author}{Braghin, F.}, \bibinfo{author}{Ruzzene, M.}, \bibinfo{year}{2020}.
\newblock \bibinfo{title}{Experimental observation of nonreciprocal band gaps in a space-time-modulated beam using a shunted piezoelectric array}.
\newblock \bibinfo{journal}{Physical Review Applied} \bibinfo{volume}{13}, \bibinfo{pages}{031001}.
%Type = Article
\bibitem[{Tessier~Brothelande et~al.(2023)Tessier~Brothelande, Cro{\"e}nne, Allein, Vasseur, Amberg, Giraud and Dubus}]{tessier2023experimental}
\bibinfo{author}{Tessier~Brothelande, S.}, \bibinfo{author}{Cro{\"e}nne, C.}, \bibinfo{author}{Allein, F.}, \bibinfo{author}{Vasseur, J.O.}, \bibinfo{author}{Amberg, M.}, \bibinfo{author}{Giraud, F.}, \bibinfo{author}{Dubus, B.}, \bibinfo{year}{2023}.
\newblock \bibinfo{title}{Experimental evidence of nonreciprocal propagation in space-time modulated piezoelectric phononic crystals}.
\newblock \bibinfo{journal}{Applied Physics Letters} \bibinfo{volume}{123}.
%Type = Article
\bibitem[{Cassedy and Oliner(1963)}]{cassedy1963dispersion}
\bibinfo{author}{Cassedy, E.S.}, \bibinfo{author}{Oliner, A.A.}, \bibinfo{year}{1963}.
\newblock \bibinfo{title}{Dispersion relations in time-space periodic media: part {I-Stable interactions}}.
\newblock \bibinfo{journal}{Proceedings of the IEEE} \bibinfo{volume}{51}, \bibinfo{pages}{1342\--1359}.
%Type = Article
\bibitem[{Cassedy(1967)}]{cassedy1967dispersion}
\bibinfo{author}{Cassedy, E.S.}, \bibinfo{year}{1967}.
\newblock \bibinfo{title}{Dispersion relations in time-space periodic media part {II-Unstable interactions}}.
\newblock \bibinfo{journal}{Proceedings of the IEEE} \bibinfo{volume}{55}, \bibinfo{pages}{1154\--1168}.
%Type = Article
\bibitem[{Trainiti and Ruzzene(2016)}]{trainiti2016non}
\bibinfo{author}{Trainiti, G.}, \bibinfo{author}{Ruzzene, M.}, \bibinfo{year}{2016}.
\newblock \bibinfo{title}{Non-reciprocal elastic wave propagation in spatiotemporal periodic structures}.
\newblock \bibinfo{journal}{New Journal of Physics} \bibinfo{volume}{18}, \bibinfo{pages}{083047}.
%Type = Article
\bibitem[{Goldsberry et~al.(2020)Goldsberry, Wallen and Haberman}]{goldsberry2020nonreciprocal}
\bibinfo{author}{Goldsberry, B.M.}, \bibinfo{author}{Wallen, S.P.}, \bibinfo{author}{Haberman, M.R.}, \bibinfo{year}{2020}.
\newblock \bibinfo{title}{Nonreciprocal vibrations of finite elastic structures with spatiotemporally modulated material properties}.
\newblock \bibinfo{journal}{Physical Review B} \bibinfo{volume}{102}, \bibinfo{pages}{014312}.
%Type = Article
\bibitem[{Nassar et~al.(2017)Nassar, Chen, Norris and Huang}]{nassar2017non}
\bibinfo{author}{Nassar, H.}, \bibinfo{author}{Chen, H.}, \bibinfo{author}{Norris, A.}, \bibinfo{author}{Huang, G.}, \bibinfo{year}{2017}.
\newblock \bibinfo{title}{Non-reciprocal flexural wave propagation in a modulated metabeam}.
\newblock \bibinfo{journal}{Extreme Mechanics Letters} \bibinfo{volume}{15}, \bibinfo{pages}{97--102}.
%Type = Article
\bibitem[{Chen et~al.(2019)Chen, Li, Nassar, Norris, Daraio and Huang}]{chen2019nonreciprocal}
\bibinfo{author}{Chen, Y.}, \bibinfo{author}{Li, X.}, \bibinfo{author}{Nassar, H.}, \bibinfo{author}{Norris, A.N.}, \bibinfo{author}{Daraio, C.}, \bibinfo{author}{Huang, G.}, \bibinfo{year}{2019}.
\newblock \bibinfo{title}{Nonreciprocal wave propagation in a continuum-based metamaterial with space-time modulated resonators}.
\newblock \bibinfo{journal}{Physical Review Applied} \bibinfo{volume}{11}, \bibinfo{pages}{064052}.
%Type = Article
\bibitem[{Wu et~al.(2021)Wu, Chen, Nassar and Huang}]{wu2021non}
\bibinfo{author}{Wu, Q.}, \bibinfo{author}{Chen, H.}, \bibinfo{author}{Nassar, H.}, \bibinfo{author}{Huang, G.}, \bibinfo{year}{2021}.
\newblock \bibinfo{title}{Non-reciprocal rayleigh wave propagation in space--time modulated surface}.
\newblock \bibinfo{journal}{Journal of the Mechanics and Physics of Solids} \bibinfo{volume}{146}, \bibinfo{pages}{104196}.
%Type = Article
\bibitem[{Palermo et~al.(2020)Palermo, Celli, Yousefzadeh, Daraio and Marzani}]{palermo2020surface}
\bibinfo{author}{Palermo, A.}, \bibinfo{author}{Celli, P.}, \bibinfo{author}{Yousefzadeh, B.}, \bibinfo{author}{Daraio, C.}, \bibinfo{author}{Marzani, A.}, \bibinfo{year}{2020}.
\newblock \bibinfo{title}{Surface wave non-reciprocity via time-modulated metamaterials}.
\newblock \bibinfo{journal}{Journal of the Mechanics and Physics of Solids} \bibinfo{volume}{145}, \bibinfo{pages}{104181}.
%Type = Article
\bibitem[{Yi et~al.(2017)Yi, Collet and Karkar}]{yi2017frequency}
\bibinfo{author}{Yi, K.}, \bibinfo{author}{Collet, M.}, \bibinfo{author}{Karkar, S.}, \bibinfo{year}{2017}.
\newblock \bibinfo{title}{Frequency conversion induced by time-space modulated media}.
\newblock \bibinfo{journal}{Physical review B} \bibinfo{volume}{96}, \bibinfo{pages}{104110}.
%Type = Article
\bibitem[{Yi et~al.(2018)Yi, Collet and Karkar}]{yi2018reflection}
\bibinfo{author}{Yi, K.}, \bibinfo{author}{Collet, M.}, \bibinfo{author}{Karkar, S.}, \bibinfo{year}{2018}.
\newblock \bibinfo{title}{Reflection and transmission of waves incident on time-space modulated media}.
\newblock \bibinfo{journal}{Physical Review B} \bibinfo{volume}{98}, \bibinfo{pages}{054109}.
%Type = Article
\bibitem[{Ye et~al.(2025)Ye, Liu, Marzani, Riva, Palermo and Wang}]{ye2025nonreciprocal}
\bibinfo{author}{Ye, Y.}, \bibinfo{author}{Liu, C.}, \bibinfo{author}{Marzani, A.}, \bibinfo{author}{Riva, E.}, \bibinfo{author}{Palermo, A.}, \bibinfo{author}{Wang, X.}, \bibinfo{year}{2025}.
\newblock \bibinfo{title}{Nonreciprocal scattering of elastic waves at time interfaces induced by spatiotemporal modulation}.
\newblock \bibinfo{journal}{arXiv preprint arXiv:2504.19385} .
%Type = Article
\bibitem[{Jin et~al.(2024)Jin, Li, Djafari-Rouhani, Torrent, Xiang and Xuan}]{jin2024exceptional}
\bibinfo{author}{Jin, Y.}, \bibinfo{author}{Li, W.}, \bibinfo{author}{Djafari-Rouhani, B.}, \bibinfo{author}{Torrent, D.}, \bibinfo{author}{Xiang, Y.}, \bibinfo{author}{Xuan, F.Z.}, \bibinfo{year}{2024}.
\newblock \bibinfo{title}{Exceptional points in time-varying oscillators with enhanced sensing sensitivity}.
\newblock \bibinfo{journal}{Physical Review Applied} \bibinfo{volume}{22}, \bibinfo{pages}{034026}.
%Type = Article
\bibitem[{Wu and Yousefzadeh(2025)}]{wu2025parametric}
\bibinfo{author}{Wu, J.}, \bibinfo{author}{Yousefzadeh, B.}, \bibinfo{year}{2025}.
\newblock \bibinfo{title}{Parametric instability in discrete models of spatiotemporally modulated materials}.
\newblock \bibinfo{journal}{arXiv preprint arXiv:2505.22970} .

\end{thebibliography}

%\end{linenumbers}
\end{document}